# Predicting Adoption Probabilities in Social Networks

Xiao Fang*    Paul J. Hu    Zhepeng Li    Weiyu Tsai

David Eccles School of Business

University of Utah

Salt Lake City, UT 84112

{xiao.fang, paul.hu, lionel.li, weiyu.tsai}@business.utah.edu

# Predicting Adoption Probabilities in Social Networks

**Abstract**

In a social network, adoption probability refers to the probability that a social entity will adopt a product, service, or opinion in the foreseeable future. Such probabilities are central to fundamental issues in social network analysis, including the influence maximization problem. In practice, adoption probabilities have significant implications for applications ranging from social network-based target marketing to political campaigns; yet, predicting adoption probabilities has not received sufficient research attention. Building on relevant social network theories, we identify and operationalize key factors that affect adoption decisions: social influence, structural equivalence, entity similarity, and confounding factors. We then develop the locally-weighted expectation-maximization method for Naïve Bayesian learning to predict adoption probabilities on the basis of these factors. The principal challenge addressed in this study is how to predict adoption probabilities in the presence of confounding factors that are generally unobserved. Using data from two large-scale social networks, we demonstrate the effectiveness of the proposed method. The empirical results also suggest that cascade methods primarily using social influence to predict adoption probabilities offer limited predictive power, and that confounding factors are critical to adoption probability predictions.

## 1. Introduction

Fostered by the ubiquitous information technology, social networks such as those facilitated by Facebook, Twitter, electronic mail, or mobile phone (Dodds et al. 2003, Kleinberg 2008, Eagle et al. 2009) have attracted increasing attention from both academia and industry that explore how to leverage such networks for greater business and societal benefits (Domingos and Richardson 2001, Pentland 2008, Chen and Zeng 2009, Weng et al. 2010, Aral et al. 2011). A salient feature of social networks is the spread of adoption behavior (e.g., adoption of a product, service, or opinion) from one social entity to another in a social network (Kleinberg 2008). This feature is central to a wide variety of applications in business (e.g., Domingos and Richardson 2001), public health (e.g., Chen et al. 2011), and politics (e.g., Carr 2008). Predicting the probability that a social entity will adopt a product, service, or opinion in the foreseeable future, namely *adoption probability*, is critical to these applications. For business organizations, such predictions are crucial to many important applications enabled by the growing proliferation of social media, such as social network-based target marketing (Hill et al. 2006), viral marketing (Domingos and Richardson 2001), and demand prediction (Hartmann 2010, Altshuler et al. 2012).[1] Consider viral marketing as an example, which targets an initial group of consumers and exploits social networks to market to a broader population. Effective viral marketing requires optimal selection of initially targeted consumers (Dye 2000, Domingos and Richardson 2001), which in turn depends on reliable prediction of adoption probabilities, because, to compare target options and search for the optimal one, one needs to predict how likely other consumers will adopt if initially targeted consumers adopt.

---

[1] Social network-based target marketing aims at making good use of social interactions among consumers for increased sales, in addition to their characteristics (Hill et al. 2006).

While predicting adoption probabilities is critical for many social network-based applications enabled by social media, it is also an important academic research problem. Studying the spread of adoption in a social network has long been a fundamental area in social sciences, particularly social computing (Kleinberg 2007). To better understand the spread of adoption, a fundamental question is how to predict (future) adoption probabilities for individuals who has not adopted by now. Answer to this question is a precursor to solving several challenging problems in social computing, such as the influence maximization problem (Kempe et al. 2005). Solutions to the problem can provide methodological foundations to novel business applications enabled by social media such as viral marketing. Existing methods for influence maximization rely on adoption probabilities that typically are set arbitrarily or assumed as given (Kempe et al. 2005, Chen et al. 2010). In actuality, adoption probabilities often are not given; however, how to predict these probabilities has been mostly unexplored (Kleinberg 2007).

Although critical from both research and practice standpoints, predicting adoption probabilities have not received adequate investigative attention. This motivates our work, to which previous studies of cascade methods (Kempe et al. 2003, Kimura and Saito 2006, Chen et al. 2009, 2010) are relevant. Building on the independent cascade model (Kleinberg 2007), cascade methods compute a social entity's likelihood of adoption from the lens of social influence (Chen et al. 2009, 2010). Cascade methods assume that a social entity's likelihood of adoption at time $t$+1 depends on his or her neighbors who become adopter at time $t$ through their social influences (Chen et al. 2009, 2010)[2]. They further assume that each adopter neighbor of a social entity influences the entity independently (Chen et al. 2009, 2010). Concretely, at time $t$, let $v$ be a nonadopter and $U$ be $v$'s neighbors who become adopter at time $t$. With these two assumptions, the probability $p_v$ of $v$ adopting at time $t$+1 is then computed as (Chen et al. 2009, 2010)

---

[2] A social entity is an adopter if he or she has adopted a product, service, or opinion, and is a nonadopter otherwise. Social neighbors of a social entity are social entities who have direct links to the social entity. An adopter neighbor is a social neighbor who is also an adopter.

$$p_v = 1 - \prod_{u \in U}(1 - p_{u,v}),  \qquad (1)$$

where $p_{u,v}$ denotes the probability that *u* influences *v* to become an adopter. Assuming that each adopter neighbor of a social entity has an equal probability of influencing that entity to become an adopter, $p_{u,v}$ is set to 1/*k* for all $u \in U$, where *k* is the number of social neighbors of *v* (Chen et al. 2009, 2010); or it is set as a constant, such as 0.1 or 0.01 (Kempe et al. 2003, Kimura and Saito 2006).

Goyal et al. (2010) propose a method to learn influence probabilities in a social network. With the same assumptions that cascade methods normally make, the influence probability method also uses equation (1) to predict the likelihood of adoption (Goyal et al. 2010). Instead of arbitrarily setting $p_{u,v}$ in equation (1) as 1/*k* or a constant, the influence probability method learns $p_{u,v}$, namely influence probability (Goyal et al. 2010), from logs of user actions. Specifically, this method learns $p_{u,v}$ as the ratio of the number of actions propagated from *u* to *v* to the total number of actions performed by *u* (Goyal et al. 2010). A review of existing methods suggests several limitations. First, these methods often depend on one factor (i.e., social influence) for predicting an entity's likelihood of adoption. In addition to social influence, there could exist other factors that affect a social entity's adoption decision, such as structural equivalence (Burt 1987), confounding factors (Van den Bulte and Lilien 2001). Methods not considering these additional factors may not holistically reveal the adoption decision; consequently, their adoption likelihood predictions could become unreliable. Second, these methods approach a social entity's adoption likelihood from a rather confined scope, typically focusing on social neighbors of the entity. However, a social entity's adoption decision could be influenced by other adopters that are not social neighbors; e.g., through structural equivalence (Burt 1987). Third, most existing methods seem to utilize partial social network data to predict adoption probabilities, normally social network structure data such as the number of social neighbors and the number of adopter neighbors. Other social network data, including intrinsic characteristics of individual entities as well as interaction intensities or relations among entities, could be important for adoption probability predictions and therefore should be considered.

To address these limitations, we target the following research questions:

(1) What key factors underlie a social entity's adoption decision? How can these factors be operationalized with more comprehensive social network data?

(2) How can we better predict adoption probabilities by considering important factors underlying individuals' adoption decisions?

(3) Although confounding factors are generally considered to be unobserved (Aral et al. 2009, Aral 2011), we cannot ignore their significance in adoption probability predictions, because they may constitute an important force of the adoption decision. Thus, a more challenging question is how to predict adoption probabilities in the presence of unobserved confounding factors.

These questions are central to novel techniques enabled by social media, such as social network-based target marketing and demand prediction that are essential to social commerce and online gaming. To address these research questions, we develop a Bayesian learning method for predicting adoption probabilities, which represents an essential contribution of our study. The proposed method employs relevant social network theories to identify key factors underlying adoption decisions; it is developed with appropriate machine learning theories and techniques. The principal challenge addressed by our method is how to predict adoption probabilities in the presence of unobserved confounding factors. We demonstrate the effectiveness of our method with data from two large-scale social networks. Our empirical results offer two interesting observations. First, cascade methods that exclusively use social influence to predict adoption probabilities seem ineffective. Second, confounding factors appear to play a significant role in adoption probability predictions, as manifested by the substantial improvement when taking such factors into consideration. Overall, our findings suggest that individuals' adoption decisions in a social network could be influenced by several related but distinct forces, above and beyond social influence, and that adoption probabilities could be better analyzed and predicted from a holistic perspective that uses more comprehensive social network data.

The rest of the paper is organized as follows. We identify key factors underlying adoption decision and propose how to operationalize these factors with social network data in §2. The problem of predicting adoption probabilities is formulated and a Bayesian learning method is then proposed to solve the problem in §3. We evaluate the effectiveness of the proposed method with data from two large-scale social networks and report our evaluation results and observations in §4. The paper is concluded with discussions of contributions, managerial implications, and limitations in §5.

## 2. Key Factors Underlying Adoption Decision: Theoretical Foundations and Operationalization

We review relevant social network theories that point to several key factors underlying a social entity's adoption decision, and then propose ways to operationalize these factors with social network data.

### 2.1 Theoretical Foundations

The social information processing model (Salancik and Pfeffer 1978) suggests that socially communicated perceptions and beliefs can influence individuals' opinions or behaviors. Specifically, *social influences*, or the impacts created through the interactions among people in a social context (Rice et al. 1990), represent an important force affecting individuals' adoption behaviors in a social network (Ibarra and Andrews 1993, Leenders 2002, Bruyn and Lilien 2008, Shalizi and Thomas 2011). Social comparison theory (Festinger 1950) also suggests that people are motivated to evaluate their opinions and behaviors by comparing themselves with others. Furthermore, social influence network theory (Friedkin 1998) posits that a person endowed with an initial opinion or behavioral assessment receives and responds to information propagated in a social network and could choose to modify an original opinion or assessment accordingly. These theories converge regarding the central role of the process by which people are influenced in communications and interactions with others, which leads to the creation of social influence (Leenders 2002). By communicating and interacting with one another, people create social influences that affect their opinions, attitudes, and behaviors (Rice et al. 1990, Leenders 2002, Iyengar et al. 2011). Bruyn and Lilien (2008) report that personal interactions among acquaintances impact not only consumption choice and purchase decision but also expectation, pre-usage attitude, and post-usage perception. People's social ties, formed through communications and interactions, enable them to learn and reflect on others' choices or opinions (Wellman 1997); the intensity of the resulting social influence reflects the strength of the social ties that connect them. In general, strong social ties entail substantial investments of time and reciprocity; therefore, they are more likely to trust each other for information sharing, opinion assessment, and decision making (Wellman 1997). As a result, people connected by stronger ties have greater influences on one another than those connected by weaker ties (Levy 1992, Levy and Nail 1993).

The structural characteristics of a social network also may affect opinions and behaviors (Burt 1987, Wejnert 2002). Network theorists argue that people develop similar opinions and behaviors through relationship patterns in a social network (Wellman 1983, Wejnert 2002). *Structural equivalence* is a fundamental structural characteristic of social network (Burt 1987); two social entities are structurally equivalent if they connect to other entities identically (Lorrain and White 1971, Wasserman and Faust 1994). Structurally equivalent people occupy the same position in the social structure and are proximate to varying extents. Such structural equivalence may be crucial to social contagion; two people identically positioned in the flow of influential communications can use the other as a frame of reference for subjective judgments and they are likely to make similar judgments, even without direct communications (Burt 1987). Therefore, weighting social entities by both structural equivalence and communications could produce more accurate adoption predictions than focusing only on their direct communications (Wejnert 2002).

Structural equivalence also moderates adoptions by affecting the homogeneity of adopters' behaviors (Wejnert 2002). Thus, people with similar social ties in a social network exhibit similar opinions or behaviors. Structural equivalence can reduce the uncertainty associated with the focal adoption, as perceived by individual entities; it represents an important force on their adoption behaviors, even in the absence of social ties that connect them directly. As Burt (1987) has noted, the spread of an opinion or behavior in a social network probably is contingent on the way the structure of the network brings people together. If they connect to the same group of people, two social entities likely exhibit similarity, because they vicariously experience or even mimic each other through the others with which they interact in the network, and thus uncertainty associated with adoption decreases (Rice and Aydin 1991). In this light, people with highly comparable social ties exhibit similar opinions or behaviors.

In addition, *entity similarity*, or the degree to which two entities in a social network are similar demographically and behaviorally (Jackson 2008), constitutes another important force that influences adoption behaviors (Salancik and Pfeffer 1978). According to Lazarsfeld and Merton (1954), major sociodemographic dimensions can stratify society-ascribed characteristics (e.g., ethnicity, gender, age) and acquired characteristics (e.g., education, occupation, behavior patterns). People with similar characteristics are more likely to exhibit similarity in their opinions and behaviors than otherwise (Aral et al. 2009). Such similarity also implies common interests and worldviews; people with highly similar demographic characteristics thus exhibit similar needs, wants, preferences, or tastes (Ibarra 1992). Entity similarity may also entail a behavioral dimension, such that people with similar behavioral profiles express similar opinions and prefer similar behaviors toward a new product or service (Ibarra 1992,

Centola 2011). All else being equal, people who have responded to products (services) similarly before are more likely to continue exhibiting that similarity in the future. Brown and Reingen (1987) and Centola (2011) also consider similarity in demographic characteristics. Finally, entity similarity can affect social entities' adoption probabilities, independent of their direct interactions (McPherson et al. 2001). The similarity of two social entities, measured demographically and behaviorally, thus could explain their adoption behaviors, in that the more similar they are, the greater similarity they exhibit in their adoption behaviors.

Adoption decisions in a social network may depend on factors beyond social influences, structural equivalence, and entity similarity though. These normally unobserved *confounding factors* appear influential in previous research (Van den Bulte and Lilien 2001, Aral et al. 2009). Aral (2011) identifies several sources of bias in both cross-sectional and longitudinal data that can confound the outcomes of social influence analyses, related to interactions and outcomes among peers, such as contextual and correlated effects (Manski 1993), unobserved heterogeneity (Van den Bulte and Lilien 2001); and marketing efforts, whose omission Iyengar et al. (2011) recognize creates upward biases in social contagion estimation. Despite the identification of several sources of confounding effects, many unobserved factors remain (Aral 2011) and may account for a significant portion of the variance in adoption behaviors observed in a social network. Our review of relevant theories thus sheds light on several important factors underlying adoption decision: social influence, structural equivalence, entity similarity, and unobserved confounding factors. We operationalize each, except unobserved confounding factors, with social network data.

**2.2 Operationalization**

Let $V = \{v_1, v_2, \cdots, v_n\}$ be a set of social entities. Pairs of social entities are linked by social ties, which can be directional or nondirectional (Wasserman and Faust 1994). For example, a social network on Twitter consists of Twitterers (i.e., social entities) connected by directional social ties (i.e., one Twitterer following another) while a social network facilitated by mobile phone service consists of service users (i.e., social entities) connected by nondirectional social ties (i.e., two-way phone communications). The strength of a social tie reflects the intensity of actions through the tie (Brown and Reingen 1987). Considering the dynamic nature of social tie strength (Kossinets and Watts 2006), we denote $x_{ij}^t$ as the strength of the social tie from social entity $v_i \in V$ to entity $v_j \in V$ at time *t*. $x_{ij}^t$ generally differs from $x_{ji}^t$ for directional ties; $x_{ij}^t$ equals $x_{ji}^t$ for nondirectional ties; and $x_{ij}^t = 0$ if there is no social tie from $v_i$ to $v_j$ until time $t$. The strength $x_{ij}^t$ of a social tie can be gauged as the aggregated intensity of actions through the tie by time $t$ (Kossinets and Watts 2006). Using the abovementioned mobile phone social network as an example, $x_{ij}^t$ can be measured as the average communication time between $v_i$ and $v_j$ by time $t$ and $x_{ij}^t = 0$ if there is no communication between the entities until time $t$.

We measure social entity $v_i's$ power of social influence (hereafter *influence power*) on entity $v_j$ at time $t$, $I_{ij}^t$, using the strength $x_{ij}^t$ of the social tie from $v_i$ to $v_j$ and we have,

$$I_{ij}^t = \frac{x_{ij}^t - x_{min}}{x_{max} - x_{min}}, \tag{2}$$

where $x_{max}$ and $x_{min}$ denote the maximum and the minimum social tie strength respectively and normalization helps avoid the dependence of $I_{ij}^t$ on the measurement unit of $x_{ij}^t$ (Han and Kamber 2006). Equation (2) is congruent with social influence theories: The stronger the tie from $v_i$ to $v_j$, the more powerful $v_i's$ influence on $v_j's$ adoption decision (Levy 1992, Levy and Nail 1993).

Social entities $v_i \in V$ and $v_j \in V$ are structurally equivalent if the following condition is satisfied (Wasserman and Faust 1994): For each $v_z \in V\setminus\{v_i, v_j\}$, whenever there is a social tie from $v_i$ to $v_z$ there is also a social tie from $v_j$ to $v_z$; moreover, whenever there is a social tie from $v_z$ to $v_i$ there is also a social tie from $v_z$ to $v_j$. Perfect structural equivalence is rare in real-world social networks (Wasserman and Faust 1994). Therefore, structural equivalence between social entities is measured as the extent to which they are structurally equivalent (Wasserman and Faust 1994). A common measure of structural equivalence is the Euclidean distance measure (Burt 1976). Let $y_{ij}^t$ be the Euclidean distance of structural equivalence between social entities $v_i$ and $v_j$ at time $t$. According to (Burt 1976, Wasserman and Faust 1994), for a social network with directional social ties, $y_{ij}^t$ is calculated as,

$$y_{ij}^t = \sqrt{\sum_{v_z \in V\setminus\{v_i,v_j\}} \left[ \left(l_{iz}^t - l_{jz}^t\right)^2 + \left(l_{zi}^t - l_{zj}^t\right)^2 \right]}; \tag{3}$$

whereas for a social network with nondirectional social ties, $y_{ij}^t$ is evaluated as,

$$y_{ij}^t = \sqrt{\sum_{v_z \in V\setminus\{v_i,v_j\}} \left(l_{iz}^t - l_{jz}^t\right)^2}, \tag{4}$$

where $l_{ab}^t = 1$ if there is a social tie from $v_a$ to $v_b$ by time $t$ and $l_{ab}^t = 0$ otherwise. Appendix A1 illustrates the calculation of $y_{ij}^t$. The higher the value of $y_{ij}^t$ the less structural equivalence between $v_i$ and $v_j$. Accordingly, we evaluate social entity $v_i$'s power of structural equivalence (hereafter *equivalence power*) on entity $v_j$ at time *t*, $E_{ij}^t$, as the following:

$$E_{ij}^t = \frac{y_{max} - y_{ij}^t}{y_{max} - y_{min}}, \tag{5}$$

where $y_{max}$ and $y_{min}$ denote the maximum and the minimum Euclidean distance of structural equivalence respectively. According to (5), the higher the structural equivalence between $v_i$ and $v_j$ (i.e., lower $y_{ij}^t$), the more powerful $v_i$'s impact on $v_j$'s adoption decision (through $v_i$'s equivalence power on $v_j$), consistent with structural equivalence theories (Burt 1987).

Entity similarity can be assessed with the distance between entity characteristics (Hand et al. 2001). A social entity is described by its intrinsic characteristics, which include time-invariant characteristics such as gender and time-variant characteristics such as behavioral characteristics. We thus represent the intrinsic characteristics of a social entity $v_i$ at time *t* using a time-dependent vector $\boldsymbol{c}_i^t$. Let $d_{ij}^t$ be the distance between intrinsic characteristics of $v_i$ and those of $v_j$ at time *t*. According to (Hand et al. 2001), we have

$$d_{ij}^t = \text{dis}(\boldsymbol{c}_i^t, \boldsymbol{c}_j^t), \tag{6}$$

where $\boldsymbol{c}_i^t$ and $\boldsymbol{c}_j^t$ are the respective characteristic vector of $v_i$ and $v_j$ at time *t*, dis(.) is a distance function, and $d_{ij}^t \geq 0$. Intrinsic characteristics of social entities differ with applications. As a result, choice of distance function is application specific (Crandall et al. 2008), as some functions are appropriate for real-valued characteristics and others are suitable for a mix of nominal and real-valued characteristics (Han and Kamber 2006). We describe the distance function used in our study in §4. $d_{ij}^t$ measures dissimilarity between $v_i$ and $v_j$. Hence, the higher the value of $d_{ij}^t$ the less entity similarity between $v_i$ and $v_j$. Like equivalence power, social entity $v_i$'s *similarity power* on entity $v_j$ at time *t*, $S_{ij}^t$, is measured as,

$$S_{ij}^t = \frac{d_{max} - d_{ij}^t}{d_{max} - d_{min}}, \tag{7}$$

where $d_{max}$ and $d_{min}$ denote the maximum and the minimum dissimilarity between social entities respectively. Congruent with theories concerning entity similarity (Ibarra 1992, Centola 2011), we formalize, in equation (7), that the higher the entity similarity between $v_i$ and $v_j$, the more powerful $v_i$'s impact on $v_j$'s adoption decision, through $v_i$'s similarity power on $v_j$. Because confounding factors are generally unobserved, their power on adoption decision (hereafter *confounding power*) is hidden. However, considering hidden confounding power is crucial, which makes it a principal challenge for predicting adoption probabilities.

## 3. Predicting Adoption Probabilities: Problem Formulation and Proposed Method

We study the problem of predicting adoption probabilities in the context that adoption of an item is diffused among social entities $V$ in a social network over a time horizon $t = 0,1,2,...$ with 0 being the start of the time horizon. The adopted item (e.g., a particular product) and the unit of time (e.g., week) are application-dependent. Under this context, the problem is defined as follows.

> Given adoption information observed at current time $T$, which includes
> 
> (1) adoption decision (i.e., adoption or not) for each social entity in $V$ by $T$;
> 
> (2) adoption time for each social entity who has adopted by $T$;
> 
> predict the probability of adopting at time $T + 1$ for each social entity who has *not* adopted until $T$.

To solve the problem, in §3.1, we analyze powers underlying a social entity's adoption decision based on observed adoption information. A method is then proposed in §3.2 to predict adoption probabilities.

### 3.1 Analysis of Powers Underlying Adoption Decision

Let $V_A^t$ denote the set of social entities who have adopted by time $t$ and $V_N^t$ represent the set of social entities who have *not* adopted until time $t$. Formally, $V_A^t = \{v_m | v_m \in V, \tau_m \leq t\}$ and $V_N^t = V \backslash V_A^t$, where $\tau_m$ is the adoption time of $v_m$. According to prevalent diffusion models in social networks (Granovetter 1978, Kleinberg 2007), a social entity's adoption decision is affected by other social entities who have already adopted. Hence, a social entity $v_j$'s adoption decision is affected by social entities in $V_A^t$ through their influence, equivalence, and similarity powers on $v_j$, where $v_j \in V_N^t$. We thus have,

$$I_j^t = \sum_{v_i \in V_A^t} I_{ij}^t, \tag{8}$$

$$E_j^t = \sum_{v_i \in V_A^t} E_{ij}^t, \tag{9}$$

$$S_j^t = \sum_{v_i \in V_A^t} S_{ij}^t, \tag{10}$$

where $I_j^t \geq 0, E_j^t \geq 0$, and $S_j^t \geq 0$ denote the total influence, total equivalence, and total similarity power on $v_j$ at time $t$ respectively; $I_{ij}^t$, $E_{ij}^t$, and $S_{ij}^t$ are given in equations (2), (5), and (7) respectively.

Receiving the powers defined in equations (8)-(10) and the hidden confounding power at time $t$, $v_j$ may become an adopter or still be an nonadopter at time $t + 1$.

---

**Construct_train** $(T, V_A^T, \{\tau_k\}, V_N^T)$
    $T$: current time
    $V_A^T$: set of adopters by $T$
    $\tau_k$: adoption time of $v_k \in V_A^T$
    $V_N^T$: set of nonadopters by $T$

//Initialization
$clock = 0$.  // $clock$ : clock time evolving from 0 to $T$
$V_{clk} = \phi$.  // $V_{clk}$ : set of adopters with adoption time equals to $clock$
$V_{early} = \phi$.  // $V_{early}$ : set of adopters with adoption time less than $clock$
$TRAIN = \phi$. // $TRAIN$ : training data
Sort social entities in $V_A^T$ by their adoption time from the earliest to the latest.

**For** each $v_k \in V_A^T$   // from the earliest adopter to the latest adopter
  **If** $(\tau_k > clock)$
    $V_{early} = V_{early} \cup V_{clk}$.
    $V_{clk} = \phi$.
    $clock = \tau_k$.
  **End if**
  $V_{clk} = V_{clk} \cup \{v_k\}$.
  $I_k^{\tau_k-1} = \sum_{v_i \in V_{early}} I_{ik}^{\tau_k-1}$.   // by (8), $I_k^{\tau_k-1}$ : total influence power on $v_k$ at time $\tau_k - 1$
  $E_k^{\tau_k-1} = \sum_{v_i \in V_{early}} E_{ik}^{\tau_k-1}$.   // by (9), $E_k^{\tau_k-1}$ : total equivalence power on $v_k$ at time $\tau_k - 1$
  $S_k^{\tau_k-1} = \sum_{v_i \in V_{early}} S_{ik}^{\tau_k-1}$.   // by (10), $S_k^{\tau_k-1}$ : total similarity power on $v_k$ at time $\tau_k - 1$
  $A_k^{\tau_k} = 1$.   //$A_k^{\tau_k}$ : $v_k's$ adoption decision at $\tau_k$ (1 – adoption; 0 – non-adoption)
  $TRAIN = TRAIN \cup \{< I_k^{\tau_k-1}, E_k^{\tau_k-1}, S_k^{\tau_k-1}, A_k^{\tau_k} >\}$ .
**End for**

**For** each $v_j \in V_N^T$
  $I_j^{T-1} = \sum_{v_i \in V_{early}} I_{ij}^{T-1}$.   // by (8), $I_j^{T-1}$ : total influence power on $v_j$ at time $T - 1$
  $E_j^{T-1} = \sum_{v_i \in V_{early}} E_{ij}^{T-1}$.   // by (9), $E_j^{T-1}$ : total equivalence power on $v_j$ at time $T - 1$
  $S_j^{T-1} = \sum_{v_i \in V_{early}} S_{ij}^{T-1}$.   // by (10), $S_j^{T-1}$ : total similarity power on $v_j$ at time $T - 1$
  $A_j^T = 0$.   //$A_j^T$ : $v_j's$ adoption decision at $T$ (1 – adoption; 0 – non-adoption)
  $TRAIN = TRAIN \cup \{< I_j^{T-1}, E_j^{T-1}, S_j^{T-1}, A_j^T >\}$ .
**End for**
**Return** $TRAIN$.

**Figure 1**: Constructing Training Data from Observed Adoption Information

We are ready to construct training data for learning adoption probabilities by analyzing the various powers from observed adoption information. At current time $T$, we observe the set $V_A^T$ of social entities who have adopted by $T$ and their adoption times as well as the set $V_N^T$ of social entities who have *not* adopted until $T$. $V_N^T$ consists of all social entities that have not adopted until $T$; that is $V_N^T = V \backslash V_A^T$. The algorithm shown in Figure 1 is developed to construct training data from observed adoption information. Two for-loops constitute the algorithm. The first for-loop computes, for each adopter in $V_A^T$, the total influence, equivalence, and similarity power that affect the adopter's decision of adoption. The second for-loop calculates, for each nonadopter in $V_N^T$, the total influence, equivalence, and similarity power received by the nonadopter at time $T - 1$, which affect the nonadopter's decision at time $T$ (i.e., the latest observable decision of the nonadopter). An illustration of the algorithm is provided in Appendix A2.

**3.2 Predicting Adoption Probabilities**

Using the algorithm shown in Figure 1, we construct training data *TRAIN* from adoption information observed at current time $T$. For a social entity $v_q$, who has *not* adopted until $T$, influence $I_q$, equivalence $E_q$, and similarity $S_q$ power on $v_q$ at time $T$ can be calculated according to equations (8)-(10):

$$I_q = \sum_{v_i \in V_A^T} I_{iq}^T,$$

$$E_q = \sum_{v_i \in V_A^T} E_{iq}^T,$$

$$S_q = \sum_{v_i \in V_A^T} S_{iq}^T.$$

Our objective is to learn from *TRAIN* the probability $P(A_q = 1 | I_q, E_q, S_q, H_q)$ of $v_q$ adopting at time $T + 1$ (i.e., $A_q = 1$) conditioning on influence $I_q$, equivalence $E_q$, similarity $S_q$ and hidden confounding $H_q$ power on $v_q$ at time $T$.[3] To achieve this objective, as illustrated in Figure 2, we face two major difficulties: (1) each record of *TRAIN* only consists of influence (*I*), equivalence (*E*), similarity (*S*) power and adoption decision (*A*) but missing hidden confounding power (*H*) and hence *TRAIN* is incomplete; (2) hidden confounding power $H_q$ on $v_q$ is also missing. In this subsection, we discuss how we tackle these difficulties.

---

[3] For simplicity, we omit time superscript for $I_q, E_q, S_q, H_q$, and $A_q$.

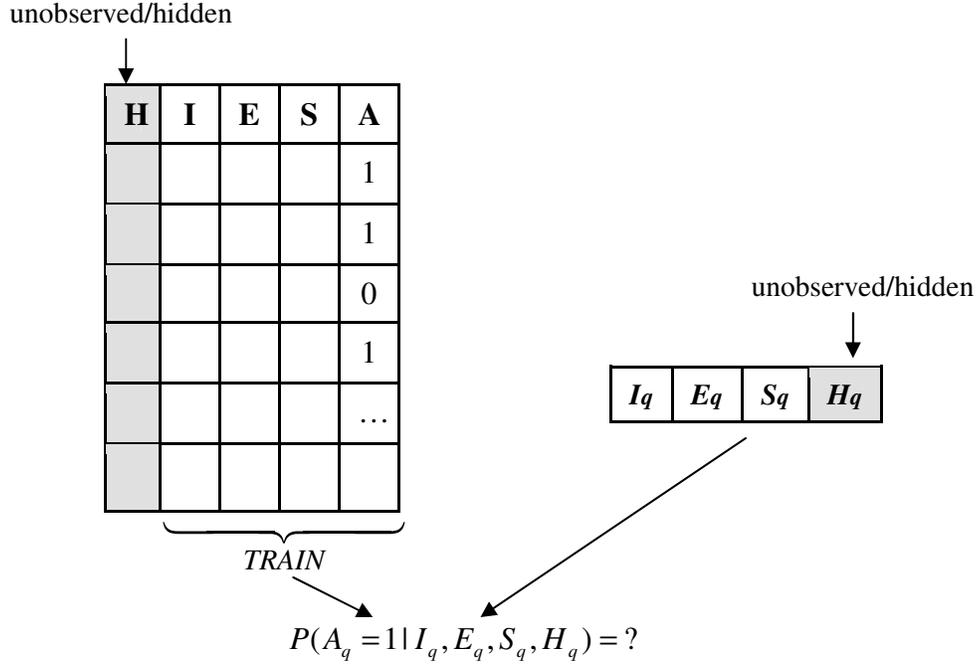

**Figure 2**: Predicting Adoption Probability $P(A_q = 1|I_q, E_q, S_q, H_q)$

Applying Bayes rule, we have

$$P(A_q = 1|I_q, E_q, S_q, H_q) = \frac{P(A_q = 1)P(I_q, E_q, S_q, H_q|A_q = 1)}{\sum_{a=0,1} P(A_q = a)P(I_q, E_q, S_q, H_q|A_q = a)}. \quad (11)$$

To compute $P(A_q = 1|I_q, E_q, S_q, H_q)$, each component in the right hand side of (11) needs to be learned from *TRAIN*. Suppose we have data on confounding power, a natural way to learn these components is the Naïve Bayes method (Mitchell 1997), which has been shown several attractive properties such as computational efficiency and good classification performance (Domingos and Pazzani 1997, Friedman 1997, Hastie et al. 2001). Note that we face the reality of not having data on confounding power in §3.2.1. The Naïve Bayes method makes a conditional independence assumption (Mitchell 1997); in our case, the method assumes that $I_q$, $E_q$, $S_q$, and $H_q$ are independent given $A_q$ and we thus obtain

$P(A_q = 1|I_q, E_q, S_q, H_q)$

$$= \frac{P(A_q = 1)P(I_q|A_q = 1)P(E_q|A_q = 1)P(S_q|A_q = 1)P(H_q|A_q = 1)}{\sum_{a=0,1} P(A_q = a)P(I_q|A_q = a)P(E_q|A_q = a)P(S_q|A_q = a)P(H_q|A_q = a)}. \quad (12)$$

Let us first consider how to estimate $P(I_q|A_q = a)$ in (12), where $a = 0,1$. For Naïve Bayes learning, representing a continuous input with an exponential family distribution (e.g., normal or exponential distribution) is a common approach (Mitchell 1997). Because influence power $I$ is continuous-valued and $I \geq 0$, we assume an exponential distribution for $I$ given adoption decision $a$ and estimate $P(I_q|A_q = a)$ as the density at $I_q$ (Mitchell 1997). That is, given $a$, $I$ follows an exponential distribution with density $f(x) = \lambda_{I|a} \exp(-\lambda_{I|a} x)$, $x \geq 0$, where $\lambda_{I|a}$ is the density parameter and $a = 0,1$. $P(I_q|A_q = a)$ can then be estimated as $f(I_q)$. We can estimate $P(E_q|A_q = a)$, $P(S_q|A_q = a)$, and $P(H_q|A_q = a)$ in the same way. Therefore, to compute $P(A_q = 1|I_q, E_q, S_q, H_q)$, we need to learn from *TRAIN* the following vector $\boldsymbol{\theta}$ of parameters:

$$\boldsymbol{\theta} = <p_1, \lambda_{I|1}, \lambda_{E|1}, \lambda_{S|1}, \lambda_{H|1}, p_0, \lambda_{I|0}, \lambda_{E|0}, \lambda_{S|0}, \lambda_{H|0}>$$

where $p_a$ is the estimate of prior probability $P(A_q = a)$ and $\lambda_{I|a}, \lambda_{E|a}, \lambda_{S|a}, \lambda_{H|a}$ are respective density parameter for influence power $I$, equivalence power $E$, similarity power $S$, hidden confounding power $H$ given adoption decision $a$, $a = 0,1$. In §3.2.1, we show how to learn $\boldsymbol{\theta}$ by addressing the difficulty of incomplete *TRAIN*; we discuss in §3.2.2 how to inference adoption probability $P(A_q = 1|I_q, E_q, S_q, H_q)$ by tackling the difficulty of hidden confounding power $H_q$.

### 3.2.1 Learning $\boldsymbol{\theta}$

Two obstacles arise when computing $P(A_q = 1|I_q, E_q, S_q, H_q)$ using (12). One is associated with learning $\boldsymbol{\theta}$. Since confounding power is hidden, we have no data in *TRAIN* for learning $\lambda_{H|a}$ directly. How can we learn $\boldsymbol{\theta}$ especially $\lambda_{H|a}$ from incomplete *TRAIN*? The second is related to the strong assumption of conditional independence by the Naïve Bayes method. How can we preserve the nice properties of the Naïve Bayes method such as computational efficiency while at the same time relax its strong assumption of conditional independence? To overcome these obstacles, we propose a method based on the classical expectation-maximization (EM) framework (Dempster et al. 1977) and the local learning theory (Atkeson et al. 1997), which are described in the following. The EM framework, developed by Dempster et al. (1977), is a widely used framework for learning from incomplete data (Bishop 2006). It is an iterative procedure starting from an initial paramter estimation (Bishop 2006). Each iteration of EM consists of the expectation step based on current paramete estimation and the maximization step, which maximizes the expectation and computes a revised parameter estimation (Bishop 2006). While regular learning employs all training data and treats each record of training data indifferently, local learning focuses on the region of training data close to the test record and weights each record of training data according to its distance to the test record (Atkeson et al. 1997). It has been shown that violations of the conditional independence assumption could be mitigated by focusing on the region of training data close to the test record (Frank et al. 2003). It is therefore appealing to integrate local learning with the Naïve Bayes method, thereby preserving the method's nice properties while relaxing the method's strong assumption to some extent.

While our method is built upon the EM framework and the local learning theory, there are challenges for which prior EM and local learning methods cannot conquer the two obstacles in our study. First, as Bishop (2006) points out, EM is a framework and its application to a problem under analysis requires problem-specific details to be defined and solved. To apply EM to our problem, we need to define the objective function to maximize and figure out how to maximize the objective function. Second, existing local learning methods are primarily developed for learning problems with observed variables. Thus, hidden confounding power poses a unique challenge for our problem: how to develop a locally weighted Naïve Bayes method in the presence of a hidden variable? Finally, how to integrate EM and local learning techniques to address both obstacles in one method? In response, we propose the locally-weighted EM method for Naïve Bayes learning.

Our proposed method learns $\boldsymbol{\theta}$ using maximum likelihood estimation. Let a record in *TRAIN* be $<I_i, E_i, S_i, A_i>$, where $i = 1,2,\cdots,n$ and $n$ is the number of records in *TRAIN*. We denote hidden confounding power for record $i$ as $H_i$. Let $D$ be complete training data and its record is $D_i = <I_i, E_i, S_i, H_i, A_i>$, where $i = 1,2,\cdots,n$. We denote $P(D|\boldsymbol{\theta})$ as the likelihood of $D$ given $\boldsymbol{\theta}$. According to (Mitchell 1997), we have,

$$P(D|\boldsymbol{\theta}) = \prod_{i=1}^{n} P(D_i|\boldsymbol{\theta}).$$

Maximum likelihood parameter estimate $\boldsymbol{\theta}_{ML}$ maximizes $P(D|\boldsymbol{\theta})$ (Mitchell 1997). It is typical to maximize $\ln[P(D|\boldsymbol{\theta})]$ instead of $P(D|\boldsymbol{\theta})$ because it is normally easier to maximize the former than the latter and parameter estimate maximizing the former also maximizes the latter (Mitchell 1997). We thus have,

$$\boldsymbol{\theta}_{ML} = argmax_{\boldsymbol{\theta}} \sum_{i=1}^{n} \ln[P(D_i|\boldsymbol{\theta})]. \tag{13}$$

To mitigate violations of the conditional independency assumption, our method weights records in $D$ differently. Let the weight of record $D_i$ be $W_i$. Following the local learning theory (Atkeson et al. 1997), we set $W_i$ higher if the distance between $D_i$ and the test record $<I_q, E_q, S_q, H_q>$ is smaller. Hence, we have,

$$W_i = K - \underbrace{[(I_i - I_q)^2 + (E_i - E_q)^2 + (S_i - S_q)^2 + (H_i - H_q)^2]}_{Square\ of\ the\ Euclidian\ Distance}, \tag{14}$$

where the distance between $D_i$ and the test record is measured as the square of the Euclidian distance between $D_i$ and the test record and $K$ is a constant. We note that both $H_i$ and $H_q$ in (14) are hidden. Having introduced weight $W_i$, our objective becomes maximizing the weighted likelihood of $D$. That is, we want to find weighted maximum likelihood parameter estimate $\boldsymbol{\theta}_{WML}$ such that

$$\boldsymbol{\theta}_{WML} = argmax_{\boldsymbol{\theta}} \sum_{i=1}^{n} W_i \ln[P(D_i|\boldsymbol{\theta})]. \tag{15}$$

It is difficult to maximize (15) directly because hidden variables exist in both $W_i$ and $P(D_i|\boldsymbol{\theta})$. Following the EM framework (Dempster et al. 1977, Bishop 2006), instead of maximizing the weighted likelihood of $D$, our method maximizes the expected weighted likelihood of $D$ given current parameter estimate $\overline{\boldsymbol{\theta}}$, where

$$\overline{\boldsymbol{\theta}} = <\bar{p}_1, \bar{\lambda}_{I|1}, \bar{\lambda}_{E|1}, \bar{\lambda}_{S|1}, \bar{\lambda}_{H|1}, \bar{p}_0, \bar{\lambda}_{I|0}, \bar{\lambda}_{E|0}, \bar{\lambda}_{S|0}, \bar{\lambda}_{H|0}>.$$

We will discuss how to set $\overline{\boldsymbol{\theta}}$ after Theorem 2. Let $f(H_q|\overline{\boldsymbol{\theta}})$ be probability density function of $H_q$ given $\overline{\boldsymbol{\theta}}$ and $f(H_i|A_i, \overline{\boldsymbol{\theta}})$ be probability density function of $H_i$ given $A_i$ and $\overline{\boldsymbol{\theta}}$. Given $\overline{\boldsymbol{\theta}}$, the expected weighted likelihood of $D$, expected on hidden variables $H_i$ and $H_q$, is expressed as,

$E_{H_i, H_q|\overline{\boldsymbol{\theta}}}\{\sum_{i=1}^{n} W_i \ln[P(D_i|\boldsymbol{\theta})]\} =$

$$\sum_{i=1}^{n} \iint \underbrace{\left[K - C_i - (H_i - H_q)^2\right]}_{W_i} \ln[P(D_i|\boldsymbol{\theta})] f(H_i|A_i, \overline{\boldsymbol{\theta}}) f(H_q|\overline{\boldsymbol{\theta}}) dH_i dH_q, \tag{16}$$

where expected weight $E_{H_i, H_q|\overline{\boldsymbol{\theta}}}[W_i] > 0$ for all $i = 1, 2, \cdots, n$,

$$E_{H_i, H_q|\overline{\boldsymbol{\theta}}}[W_i] = \iint \left[K - C_i - (H_i - H_q)^2\right] f(H_i|A_i, \overline{\boldsymbol{\theta}}) f(H_q|\overline{\boldsymbol{\theta}}) dH_i dH_q, \tag{17}$$

$$C_i = (I_i - I_q)^2 + (E_i - E_q)^2 + (S_i - S_q)^2. \tag{18}$$

We set $K$ adequately such that $E_{H_i, H_q|\overline{\boldsymbol{\theta}}}[W_i] > 0$ for all $i = 1, 2, \cdots, n$, and we set

$$K = C_i^{max} + \frac{2}{\bar{\lambda}_{H|0}^2} + \frac{2}{\bar{\lambda}_{H|1}^2}, \tag{19}$$

where $C_i^{max}$ represents the maximum among all $C_i$, $i = 1, 2, \cdots, n$, $\bar{\lambda}_{H|0}^2$ and $\bar{\lambda}_{H|1}^2$ denote the square of $\bar{\lambda}_{H|0}$ and $\bar{\lambda}_{H|1}$ respectively. Derivation of (19) is given in Appendix B.

We want to find $\hat{\boldsymbol{\theta}}$ that maximizes the expected weighted likelihood of $D$ expressed in (16). That is,

$$\hat{\boldsymbol{\theta}} = argmax_{\boldsymbol{\theta}} E_{H_i, H_q | \bar{\boldsymbol{\theta}}} \{\sum_{i=1}^{n} W_i \ln[P(D_i|\boldsymbol{\theta})]\}, \tag{20}$$

where

$$\hat{\boldsymbol{\theta}} = <\hat{p}_1, \hat{\lambda}_{I|1}, \hat{\lambda}_{E|1}, \hat{\lambda}_{S|1}, \hat{\lambda}_{H|1}, \hat{p}_0, \hat{\lambda}_{I|0}, \hat{\lambda}_{E|0}, \hat{\lambda}_{S|0}, \hat{\lambda}_{H|0}>.$$

To maximize (20), we show in Theorem 1 that the Hessian matrix of $E_{H_i, H_q | \bar{\boldsymbol{\theta}}} \{\sum_{i=1}^{n} W_i \ln[P(D_i|\boldsymbol{\theta})]\}$ is negative definite, which satisfies the sufficient condition to maximize a multivariable function (Greene 2008); we further show in Theorem 2 the necessary condition (Greene 2008) that all first-order partial derivatives of $E_{H_i, H_q | \bar{\boldsymbol{\theta}}} \{\sum_{i=1}^{n} W_i \ln[P(D_i|\boldsymbol{\theta})]\}$ equal 0 is met and give closed form solution for each parameter in $\hat{\boldsymbol{\theta}}$.

**Theorem 1.** The Hessian matrix of $E_{H_i, H_q | \bar{\boldsymbol{\theta}}} \{\sum_{i=1}^{n} W_i \ln[P(D_i|\boldsymbol{\theta})]\}$ is negative definite.

Proof: See Appendix C1.

**Theorem 2.** Given $\bar{\boldsymbol{\theta}}$, parameters in $\hat{\boldsymbol{\theta}}$ that maximize the expected weighted likelihood of $D$ are derived as below.

$$\hat{p}_1 = \frac{\sum_{i=1}^{n} A_i Q_i}{\sum_{i=1}^{n} [A_i Q_i + (1 - A_i) R_i]} \tag{21}$$

$$\hat{p}_0 = 1 - \hat{p}_1 \tag{22}$$

$$\hat{\lambda}_{I|1} = \frac{\sum_{i=1}^{n} A_i Q_i}{\sum_{i=1}^{n} A_i I_i Q_i} \tag{23}$$

$$\hat{\lambda}_{I|0} = \frac{\sum_{i=1}^{n} (1 - A_i) R_i}{\sum_{i=1}^{n} (1 - A_i) I_i R_i} \tag{24}$$

$$\hat{\lambda}_{E|1} = \frac{\sum_{i=1}^{n} A_i Q_i}{\sum_{i=1}^{n} A_i E_i Q_i} \tag{25}$$

$$\hat{\lambda}_{E|0} = \frac{\sum_{i=1}^{n} (1 - A_i) R_i}{\sum_{i=1}^{n} (1 - A_i) E_i R_i} \tag{26}$$

$$\hat{\lambda}_{S|1} = \frac{\sum_{i=1}^{n} A_i Q_i}{\sum_{i=1}^{n} A_i S_i Q_i} \tag{27}$$

$$\hat{\lambda}_{S|0} = \frac{\sum_{i=1}^{n}(1-A_i)R_i}{\sum_{i=1}^{n}(1-A_i)S_iR_i} \tag{28}$$

$$\hat{\lambda}_{H|1} = \frac{\sum_{i=1}^{n}A_i\bar{\lambda}_{H|1}Q_i}{\sum_{i=1}^{n}A_i[R_i + 2\bar{\lambda}_{H|1}^2 + 2\bar{p}_1(2\bar{\lambda}_{H|0}^2 + \bar{\lambda}_{H|1}^2 - 3\bar{\lambda}_{H|0}\bar{\lambda}_{H|1}) - 6\bar{\lambda}_{H|0}^2]} \tag{29}$$

$$\hat{\lambda}_{H|0} = \frac{\sum_{i=1}^{n}(1-A_i)\bar{\lambda}_{H|0}R_i}{\sum_{i=1}^{n}(1-A_i)[Q_i + 2\bar{\lambda}_{H|0}^2 + 2\bar{p}_0(2\bar{\lambda}_{H|1}^2 + \bar{\lambda}_{H|0}^2 - 3\bar{\lambda}_{H|0}\bar{\lambda}_{H|1}) - 6\bar{\lambda}_{H|1}^2]} \tag{30}$$

In equations (21)-(30),

$$Q_i = (K - C_i)\bar{\lambda}_{H|0}^2\bar{\lambda}_{H|1}^2 - 2\bar{\lambda}_{H|0}^2 - 2\bar{p}_0(\bar{\lambda}_{H|1}^2 - \bar{\lambda}_{H|0}\bar{\lambda}_{H|1}),$$

$$R_i = (K - C_i)\bar{\lambda}_{H|0}^2\bar{\lambda}_{H|1}^2 - 2\bar{\lambda}_{H|1}^2 - 2\bar{p}_1(\bar{\lambda}_{H|0}^2 - \bar{\lambda}_{H|0}\bar{\lambda}_{H|1}).$$

Proof: See Appendix C2.

By Theorems 1 and 2, we can compute $\hat{\boldsymbol{\theta}}$ from $\bar{\boldsymbol{\theta}}$ and training data *TRAIN*. While *TRAIN* is constructed using the algorithm shown in Figure 1, we need to initialize $\bar{\boldsymbol{\theta}}$. We set $\bar{p}_1, \bar{p}_0, \bar{\lambda}_{I|1}, \bar{\lambda}_{I|0}, \bar{\lambda}_{E|1}, \bar{\lambda}_{E|0}, \bar{\lambda}_{S|1}$, and $\bar{\lambda}_{S|0}$ in $\bar{\boldsymbol{\theta}}$ as maximum likelihood estimates of these parameters from *TRAIN*. The estimates, as shown in equations (31)-(38), are standard estimates for Naïve Bayes learning and derivations of them can be found in (Mitchell 1997). It is more appropriate to learn the effect of confounding power on parameter estimates from data and adjust parameter estimates accordingly than setting the effect arbitrarily during the initialization of $\bar{\boldsymbol{\theta}}$. Thus, parameter estimates in (31)-(38) do not consider the following factors: confounding power and weight, which takes confounding power as a component term according to (14). The effect of these factors on parameter estimates will be learned and used to adjust parameter estimates when computing $\hat{\boldsymbol{\theta}}$ from $\bar{\boldsymbol{\theta}}$.

$$\bar{p}_1 = \frac{\sum_{i=1}^{n}A_i}{n} \tag{31}$$

$$\bar{p}_0 = 1 - \bar{p}_1 \tag{32}$$

$$\bar{\lambda}_{I|1} = \frac{\sum_{i=1}^{n}A_i}{\sum_{i=1}^{n}A_iI_i} \tag{33}$$

$$\bar{\lambda}_{I|0} = \frac{\sum_{i=1}^{n}(1-A_i)}{\sum_{i=1}^{n}(1-A_i)I_i} \tag{34}$$

$$\bar{\lambda}_{E|1} = \frac{\sum_{i=1}^{n}A_i}{\sum_{i=1}^{n}A_iE_i} \tag{35}$$

$$\bar{\lambda}_{E|0} = \frac{\sum_{i=1}^{n}(1-A_i)}{\sum_{i=1}^{n}(1-A_i)E_i} \tag{36}$$

$$\bar{\lambda}_{S|1} = \frac{\sum_{i=1}^{n}A_i}{\sum_{i=1}^{n}A_iS_i} \tag{37}$$

$$\bar{\lambda}_{S|0} = \frac{\sum_{i=1}^{n}(1-A_i)}{\sum_{i=1}^{n}(1-A_i)S_i} \tag{38}$$

We consider three possible cases for $\bar{\lambda}_{H|a}$: given adoption decision *a*, on average, confounding power is the largest, average, or smallest among all powers underlying adoption decision. These cases are modeled using equations (39)-(41) respectively:

$$\frac{1}{\bar{\lambda}_{H|a}} = \max\left(\frac{1}{\bar{\lambda}_{I|a}}, \frac{1}{\bar{\lambda}_{E|a}}, \frac{1}{\bar{\lambda}_{S|a}}\right) \times (1+\varepsilon_1), \quad a=0,1 \quad \text{(largest case)} \tag{39}$$

$$\frac{1}{\bar{\lambda}_{H|a}} = \frac{\left(\frac{1}{\bar{\lambda}_{I|a}} + \frac{1}{\bar{\lambda}_{E|a}} + \frac{1}{\bar{\lambda}_{S|a}}\right)}{3} \times (1+\varepsilon_2), \quad a=0,1 \quad \text{(average case)} \tag{40}$$

$$\frac{1}{\bar{\lambda}_{H|a}} = \min\left(\frac{1}{\bar{\lambda}_{I|a}}, \frac{1}{\bar{\lambda}_{E|a}}, \frac{1}{\bar{\lambda}_{S|a}}\right) \times (1-\varepsilon_3), \quad a=0,1 \quad \text{(smallest case)} \tag{41}$$

where $\frac{1}{\bar{\lambda}_{H|a}}, \frac{1}{\bar{\lambda}_{I|a}}, \frac{1}{\bar{\lambda}_{E|a}}$, and $\frac{1}{\bar{\lambda}_{S|a}}$ are the mean of confounding, influence, equivalence, and similarity power respectively. Random terms $\varepsilon_1, \varepsilon_2$ and $\varepsilon_3$ introduce randomness into the equations; in our empirical study, we set $\varepsilon_1$ and $\varepsilon_3$ uniformly distributed over (0, 0.01) and $\varepsilon_2$ uniformly distributed over (-0.005, 0.005) respectively. Rather than choosing a case arbitrarily, we consider all three cases and let data decide which case is most appropriate. Therefore, three different initializations of $\bar{\theta}$ are set: equations (31)-(38) and (39), equations (31)-(38) and (40), and equations (31)-(38) and (41).

We propose the locally-weighted EM method for Naïve Bayes learning (LEMNB) in Figure 3. The method learns parameter estimate $\hat{\theta}$ from *TRAIN* for a social entity receiving influence $I_q$, equivalence $E_q$, and similarity $S_q$ power. LEMNB first creates bootstrap samples form *TRAIN*[4]. It then loops through the three cases (for-loop on *k*). Each case consists of *N* trials (for-loop on *h*). $\bar{\theta}_{kh}$ denotes $h^{th}$ trial of parameter initialization for case *k*. For each $\overline{\boldsymbol{\theta}}_{kh}$, as illustrated in Figure 4, the method computes $\widehat{\boldsymbol{\theta}}_{kh}$ iteratively through bootstrap samples. Bootstrap sampling introduces variations into training data and thus reduces the chance of overfitting. As a byproduct of computing $\widehat{\boldsymbol{\theta}}_{kh}$, LEMNB also calculates the maximum expected weighted likelihood $E_{kh}$. For each case, average parameter estimate $\widehat{\boldsymbol{\theta}}_k$ and average likelihood $E_k$ across *N* trials are calculated. The most appropriate case is determined as the one with the largest $E_k$ and the parameter estimate $\widehat{\boldsymbol{\theta}}_k$ of this case is used as the final parameter estimate. We set the number of bootstrap samples *M* and the number of trials *N* adequately such that $\widehat{\boldsymbol{\theta}}_{kh}$ and $\widehat{\boldsymbol{\theta}}_k$ converge.

---

[4] Bootstrap sampling creates a sample, which has the same number of records as *TRAIN*, by repeatedly sampling from *TRAIN* with replacement (Bishop 2006).

**LEMNB** (*M*, *N*, $I_q, E_q, S_q$)
    *M*: number of bootstrap samples
    *N*: number of trials
    $I_q, E_q, S_q$: influence, equivalence, similarity power

Create *M* bootstrap samples from *TRAIN*: *TRAIN*₁, *TRAIN*₂, …,*TRAIN*_M .
**For** *k* = 1 to 3 step 1
    **For** *h* = 1 to *N* step 1
        *TRAIN* = *TRAIN*₁.
        Generate random term $\varepsilon_1, \varepsilon_2$ or $\varepsilon_3$.
        **If** (*k* = 1)
            Initialize $\overline{\boldsymbol{\theta}}_{kh}$ according to equations (31)-(38) and (39).
        **Else if** (*k* = 2)
            Initialize $\overline{\boldsymbol{\theta}}_{kh}$ according to equations (31)-(38) and (40).
        **Else**
            Initialize $\overline{\boldsymbol{\theta}}_{kh}$ according to equations (31)-(38) and (41).
        **End if**
        **For** *j* = 2 to *M* step 1
            *TRAIN* = *TRAIN*_j.
            Compute $\widehat{\boldsymbol{\theta}}_{kh}$ from $\overline{\boldsymbol{\theta}}_{kh}$ and *TRAIN* according to equations (21)-(30).
            $E_{kh} = E_{H_i, H_q | \overline{\boldsymbol{\theta}}_{kh}} \{\sum_{i=1}^{n} W_i \ln[P(D_i | \widehat{\boldsymbol{\theta}}_{kh})]\}$.
            **If** (*j* < *M*)
                $\overline{\boldsymbol{\theta}}_{kh} = \widehat{\boldsymbol{\theta}}_{kh}$.
            **End if**
        **End for**
    **End for**
    $\widehat{\boldsymbol{\theta}}_k = \sum_{h=1}^{N} \widehat{\boldsymbol{\theta}}_{kh} / N$.
    $E_k = \sum_{h=1}^{N} E_{kh} / N$.
**End for**
$\widehat{\boldsymbol{\theta}} = \widehat{\boldsymbol{\theta}}_k$ with the largest $E_k$ for *k* = 1,2,3.
**Return** $\widehat{\boldsymbol{\theta}}$.

**Figure 3**: The Locally-weighted EM method for Naïve Bayes Learning (LEMNB)

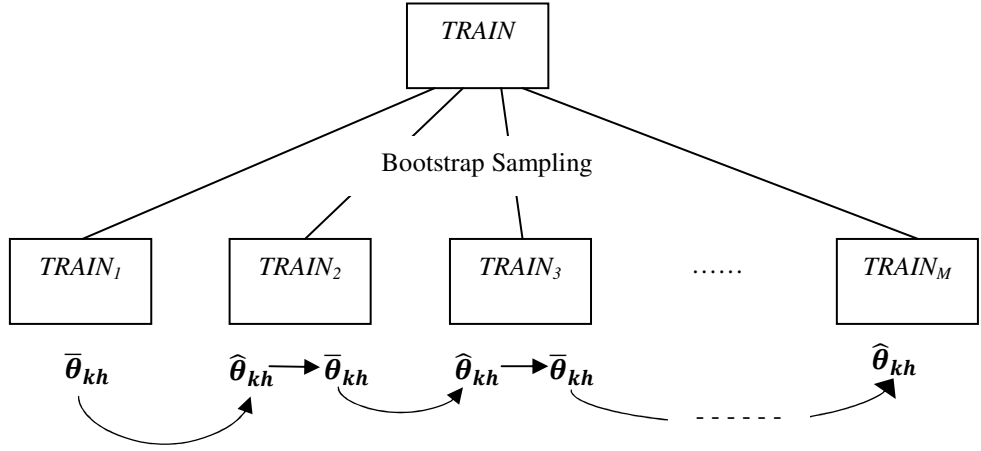

**Figure 4**: Computing $\widehat{\boldsymbol{\theta}}_{kh}$

### 3.2.2 Inference

Having learned parameter estimate $\widehat{\boldsymbol{\theta}}$, inference adoption probability is a relatively easier task. Given parameter estimate $\widehat{\boldsymbol{\theta}} = <\hat{p}_1, \hat{\lambda}_{I|1}, \hat{\lambda}_{E|1}, \hat{\lambda}_{S|1}, \hat{\lambda}_{H|1}, \hat{p}_0, \hat{\lambda}_{I|0}, \hat{\lambda}_{E|0}, \hat{\lambda}_{S|0}, \hat{\lambda}_{H|0}>$, by (12), we have

$$P(A_q = 1|I_q, E_q, S_q, H_q) = \frac{P(A_q = 1)P(I_q|A_q = 1)P(E_q|A_q = 1)P(S_q|A_q = 1)P(H_q|A_q = 1)}{\sum_{a=0,1} P(A_q = a)P(I_q|A_q = a)P(E_q|A_q = a)P(S_q|A_q = a)P(H_q|A_q = a)}$$

$$= \frac{\hat{p}_1 \hat{\lambda}_{I|1}\exp(-\hat{\lambda}_{I|1}I_q)\hat{\lambda}_{E|1}\exp(-\hat{\lambda}_{E|1}E_q)\hat{\lambda}_{S|1}\exp(-\hat{\lambda}_{S|1}S_q)\hat{\lambda}_{H|1}\exp(-\hat{\lambda}_{H|1}H_q)}{\sum_{a=0,1} \hat{p}_a \hat{\lambda}_{I|a}\exp(-\hat{\lambda}_{I|a}I_q)\hat{\lambda}_{E|a}\exp(-\hat{\lambda}_{E|a}E_q)\hat{\lambda}_{S|a}\exp(-\hat{\lambda}_{S|a}S_q)\hat{\lambda}_{H|a}\exp(-\hat{\lambda}_{H|a}H_q)}. \quad (42)$$

To compute adoption probability using (42), the only difficulty is the hidden variable $H_q$. However, given $\widehat{\boldsymbol{\theta}}$, probability density of $H_q$ is known and we have

$$f(H_q|\widehat{\boldsymbol{\theta}}) = f(H_q|A_q = 1, \widehat{\boldsymbol{\theta}})P(A_q = 1|\widehat{\boldsymbol{\theta}}) + f(H_q|A_q = 0, \widehat{\boldsymbol{\theta}})P(A_q = 0|\widehat{\boldsymbol{\theta}}). \quad (43)$$

Therefore, we can approximate adoption probability $P(A_q = 1|I_q, E_q, S_q, H_q)$ with its expectation on $H_q$ and compute the expectation using Monte Carlo method (Bishop 2006)[5]. As shown in Figure 5, the adoption probability inference algorithm repeatedly generates a sample of $H_q$ and calculates adoption probability using the sample and equation (42) until convergence. The algorithm outputs the expectation of the adoption probability, i.e., expected on $H_q$.

---

[5] There is no closed form representation of the expectation and Monte Carlo method is a viable approach (Bishop 2006).

> **Inference** ($\hat{\boldsymbol{\theta}}, I_q, E_q, S_q, \sigma$)
>     $\hat{\boldsymbol{\theta}}$: parameter estimate
>     $I_q, E_q, S_q$: influence, equivalence, similarity power
>     $\sigma$: predefined convergence threshold
>
> count = 0.
> sum = 0.
> mean = 0.
> **Do**
>     pre_mean = mean.
>     Generate a sample $h_q$ of $H_q$ according to its probability density in (43).
>     Calculate $P(A_q = 1 | I_q, E_q, S_q, h_q)$ using (42).
>     sum = sum + $P(A_q = 1 | I_q, E_q, S_q, h_q)$.
>     count = count + 1.
>     mean = sum/count.
> **Until** ($\frac{|\text{mean} - \text{pre\_mean}|}{\text{mean}} \leq \sigma$)
> **Return** mean.

**Figure 5**: Inferencing Adoption Probability

## 4. Empirical Evaluations

We evaluated the proposed method with data from two large-scale social networks. One is a social network of communications among mobile phone users. Mobile communication-based social networks have been used to evaluate theories, methods, or applications in social network research (Pentland 2008, Eagle et al. 2009, 2010). We further examined the proposed method with a virtual world social network – a social network of avatars[6]. This additional evaluation is closely related to online gaming, a fast-growing industry that has been transformed by social media (Zukerman and Albrecht 2001, Hemp 2006). The proposed method could be highly beneficial for marketing virtual items to game players, a fast-growing revenue source for online game service providers. According to PlaySpan (2012), game players in the United States purchased $2.3 billion worth of virtual items in 2011– nearly 30% growth compared with 2009. By using our method, service providers could more effectively prioritize their marketing efforts and focus on consumers more likely to adopt new virtual products in the next time period. We report evaluation results with the mobile social network in this section. We observe similar evaluation results in the avatar social network and report key findings in Appendix D for space consideration.

---

[6] An avatar is a user-created character in a virtual community.

## 4.1 Data and Evaluation Procedure

Our evaluations used data sets collected from a major mobile service provider. One data set consists of 14.7 million records of mobile phone communications among 34,797 users over a one-year period. Each record contains the identifications of the two users connected by a communication, and the time and duration of the communication.[7] Another data set contains profiles of the 34,797 users. Each user profile is comprised of 2 time-invariant demographic characteristics: gender and age, and 18 time-variant behavioral characteristics: the frequency of using each of the 18 calling modes offered by the service provider.[8] We also collected data on week-by-week adoption of a mobile service *A* launched at the beginning of the study period, i.e., who, out of the 34,797 users, adopted the service in which week.[9] Consistent with existing literature (Bass 1969, Iyengar et al. 2011), the week of adopting service *A* is the week of initial purchase of *A*. Figure 6 shows weekly adoption rates over the entire study period.[10] On average, 110 users adopted the service in a week, yielding an average weekly adoption rate of 0.42%. The unit of time in our evaluations was week because of weekly adoption data. We thus constructed from data 52 snapshots of the mobile social network, each of which corresponded to the social network by the end of week $t$, $t = 1,2,...,52$. In each snapshot, social entities ($v_i$) corresponded to users; a social tie between social entities $v_i$ and $v_j$ was created if there was communication between their corresponding users and the strength of the social tie $x_{ij}^t$ was measured as the average weekly communication time between the users by week $t$; and the intrinsic characteristics $\boldsymbol{c_i^t}$ of a social entity $v_i$ by week *t* consisted of the 2 time-invariant characteristics and the 18 time-variant characteristics.

---

[7] Real user identifications are replaced with unique numbers for privacy consideration.
[8] While frequencies of using calling modes change every day, gender and age do not change within the one-year period.
[9] Service name is not disclosed according to our agreement with the data provider.
[10] Adoption rate of a week is calculated as the ratio between the number of adopters who adopt during the week and the number of nonadopters at the beginning of the week.

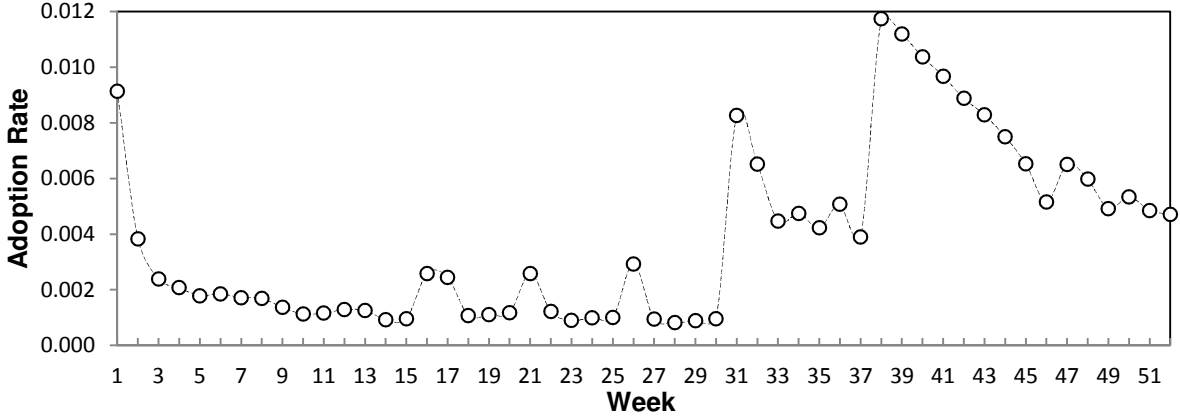

**Figure 6**: Weekly Adoption Rate

In an evaluation, we picked a week as current time $T$. Using social network data and adoption data by the end of week $T$ as training data, we applied the proposed method and eight benchmark methods to predict adoption probability in week $T + 1$ for each social entity who has not adopted by the end of week $T$. Specifically, we predicted adoption probabilities of social entities that will adopt in week $T + 1$ as well as those that will not adopt in week $T + 1$. Several implementation details of the proposed method warrant descriptions. Because intrinsic characteristics $c_i^t$ involve different types of attribute, we employed a standard distance function for measuring the similarity between entities with a mix of nominal, real-valued, and integer attributes (Han and Kamber 2006). For two entities with $n$ attributes, $x = (x_1, x_2, \ldots, x_n)$ and $y = (y_1, y_2, \ldots, y_n)$, where $x_i$ is the value of the $i$th attribute for entity $x$, $i = 1, 2, \ldots n$, the distance between $x$ and $y$ for each attribute is measured first. Let $d(x_i, y_i)$ be the distance between $x$ and $y$ for the $i$th attribute. According to (Han and Kamber 2006), if the $i$th attribute is nominal, $d(x_i, y_i)$ is given by

$$d(x_i, y_i) = \begin{cases} 0, & \text{if } x_i = y_i \\ 1; & \text{otherwise} \end{cases} \tag{44}$$

if the $i$th attribute is real-valued or integer, $d(x_i, y_i)$ is

$$d(x_i, y_i) = \frac{|x_i - y_i|}{\max_i - \min_i}, \tag{45}$$

where $\max_i$ and $\min_i$ denote the maximum value and the minimum value of the $i$th attribute among all entities, respectively. The distance between $x$ and $y$ then can be computed by integrating their distance on each attribute,

$$d(x,y) = \frac{\sum_{i=1}^{n} d(x_i, y_i)}{n}. \tag{46}$$

In equations (2), (5), and (7), $x_{min}$, $y_{min}$, and $d_{min}$ were set to 0; $x_{max}$, $y_{max}$, and $d_{max}$ were set to equal the maximum social tie strength, the maximum Euclidean distance of structural equivalence, and the maximum dissimilarity between social entities by the end of week $T$ respectively. We set $M$ and $N$ in the LEMNB method (Figure 3) to 5 and 20 respectively.

Prior studies of cascade methods are related to our work. Therefore, three cascade methods (Chen et al. 2009), namely CM1, CM2, and CM3, were benchmarked. Derived from (1), a cascade method computes the probability $p_v$ of $v$ adopting in week $T$+1 as

$$p_v = 1 - (1-p)^l, \tag{47}$$

where $l$ is the number of $v$'s neighbors who adopt in week $T$, $p$ is set to 1/$k$, 0.1, and 0.01 respectively by CM1, CM2, and CM3 (Chen et al. 2009, Kempe et al. 2003), and $k$ is the number of neighbors of $v$ in week $T$. We also compared the proposed method with the influence probability method (Goyal et al. 2010), which calculates adoption probabilities with equation (1) and learns $p_{u,v}$ in (1) as the ratio between the number of actions propagated from $u$ to $v$ and the total number of actions performed by $u$. In the context of this study, actions in (Goyal et al. 2010) refer to service adoptions. To implement the influence probability method, we gathered additional data regarding weekly adoptions of all *other* mobile services by the same group of users during the one-year study period and used the additional data to calculate $p_{u,v}$.[11] According to Goyal et al. (2010), if $v$ adopts service B after $u$'s adoption of B and there exists a social tie connecting $u$ and $v$ before $u$'s adoption of B, adoption of service B is considered to be propagated from $u$ to $v$. If $u$ adopts services B, C, and D, but only B is propagated from $u$ to $v$, then $p_{u,v} = 1/3$ (Goyal et al. 2010). We also included representative classification methods as benchmarks. Because the proposed method is built on the Naïve Bayes method and employs local learning techniques, we compared it with the Naïve Bayes method (Mitchell 1997), the locally weighted Naive Bayes method (Frank et al. 2003) and the k-nearest neighbor method (Hand et al. 2001). We also benchmarked against Support Vector Machine (Burges 1998), which has been shown good predictive power among classification methods (Huang and Ling 2005). Table 1 summarizes the methods compared in the evaluations.

---

[11] All mobile services we studied were launched during the one-year study period. None of the mobile services we studied are necessities.

| Method | Abbreviation | Note |
|---|---|---|
| locally-weighted EM method for Naïve Bayes Learning | LEMNB | Proposed method |
| cascade methods | CM1, CM2, CM3 | Benchmark |
| influence probability method | IP | Benchmark |
| Naïve Bayes | NB | Benchmark |
| locally weighted Naive Bayes | LWNB | Benchmark |
| k-nearest neighbor | k-NN | Benchmark |
| support vector machine | SVM | Benchmark |

**Table 1:** Methods Compared in the Evaluations

**4.2 Evaluation Results and Analyses**

Following the evaluation procedure, we conducted 50 evaluations to compare the proposed method and the benchmarks with $T$ ranging from 2 to 51. The performance of each method was evaluated using AUC, the area under the ROC curve (Fawcett 2006). AUC is a standard metric for assessing methods that predict classification probabilities (Huang and Ling 2005). According to Fawcett (2006), AUC is equivalent to the probability that a randomly chosen positive instance will be predicted to have a higher probability of belonging to the positive class than a randomly chosen negative instance. In our case, AUC is equivalent to the probability that a randomly chosen adopter will be predicted to have a higher adoption probability than a randomly chosen nonadopter. Therefore, a method that yields a higher AUC generally offers greater predictive power than a method that produces a lower AUC (Fawcett 2006).

In Table 2, we show AUCs of the proposed method and those of the benchmark methods across 50 evaluations. To examine whether our proposed method outperformed each benchmark method, we conducted the Wilcoxon signed-ranks test (Demsar 2006), which is widely used for comparing the performance of predictive methods on the basis of AUC (Demsar 2006). We applied the Wilcoxon test to the AUCs in Table 2; testing results suggested that the proposed method significantly outperformed each benchmark method ($p < 0.001$). In addition to demonstrating the superior predictive power of our method over each benchmark method statistically, we also observed substantial AUC improvements by our method over the benchmarks, as shown in Table 2. For example, across all 50 evaluations, the average AUC of our method is 0.8029, while that of k-NN is 0.6910.

| Evaluation Week ($T+1$) | LEMNB | CM1 | CM2 | CM3 | IP | NB | LWNB | SVM | k-NN |
|---|---|---|---|---|---|---|---|---|---|
| 3 | 0.8722 | 0.5166 | 0.5167 | 0.5167 | 0.5295 | 0.5254 | 0.7050 | 0.8257 | 0.6675 |
| 4 | 0.8299 | 0.5023 | 0.5023 | 0.5023 | 0.5196 | 0.8181 | 0.6202 | 0.7664 | 0.7113 |
| 5 | 0.8505 | 0.5019 | 0.5019 | 0.5019 | 0.5131 | 0.7809 | 0.6120 | 0.7691 | 0.6555 |
| 6 | 0.8666 | 0.5025 | 0.5025 | 0.5025 | 0.5169 | 0.8072 | 0.6839 | 0.7786 | 0.7489 |
| 7 | 0.8643 | 0.5032 | 0.5032 | 0.5032 | 0.5218 | 0.8241 | 0.6701 | 0.7319 | 0.7713 |
| 8 | 0.8286 | 0.5030 | 0.5030 | 0.5030 | 0.5258 | 0.8121 | 0.5565 | 0.6518 | 0.7516 |
| 9 | 0.8083 | 0.5023 | 0.5023 | 0.5023 | 0.5300 | 0.7200 | 0.7613 | 0.7657 | 0.7087 |
| 10 | 0.8011 | 0.5013 | 0.5013 | 0.5013 | 0.5339 | 0.7826 | 0.6340 | 0.6339 | 0.7245 |
| 11 | 0.8643 | 0.5032 | 0.5032 | 0.5032 | 0.5381 | 0.8200 | 0.7903 | 0.5123 | 0.7901 |
| 12 | 0.7974 | 0.5030 | 0.5030 | 0.5030 | 0.5313 | 0.7422 | 0.7680 | 0.5508 | 0.6769 |
| 13 | 0.8339 | 0.5021 | 0.5021 | 0.5021 | 0.5256 | 0.7244 | 0.8004 | 0.6020 | 0.6446 |
| 14 | 0.8223 | 0.5020 | 0.5020 | 0.5020 | 0.5291 | 0.7315 | 0.7962 | 0.6269 | 0.7369 |
| 15 | 0.8654 | 0.5022 | 0.5022 | 0.5022 | 0.5132 | 0.8656 | 0.6106 | 0.8112 | 0.7103 |
| 16 | 0.7240 | 0.5020 | 0.5020 | 0.5020 | 0.5292 | 0.7046 | 0.6813 | 0.6822 | 0.6718 |
| 17 | 0.7270 | 0.5033 | 0.5033 | 0.5033 | 0.5293 | 0.7460 | 0.7468 | 0.6122 | 0.6814 |
| 18 | 0.8276 | 0.5022 | 0.5022 | 0.5022 | 0.5272 | 0.7967 | 0.7389 | 0.7972 | 0.7359 |
| 19 | 0.8323 | 0.5035 | 0.5035 | 0.5035 | 0.5262 | 0.8458 | 0.7796 | 0.8342 | 0.7333 |
| 20 | 0.7221 | 0.5028 | 0.5028 | 0.5028 | 0.5103 | 0.7215 | 0.7653 | 0.6694 | 0.6439 |
| 21 | 0.7852 | 0.5050 | 0.5050 | 0.5050 | 0.5195 | 0.7967 | 0.7614 | 0.7138 | 0.6604 |
| 22 | 0.8497 | 0.5015 | 0.5015 | 0.5015 | 0.5267 | 0.8571 | 0.8049 | 0.7181 | 0.7329 |
| 23 | 0.7905 | 0.5026 | 0.5026 | 0.5026 | 0.5295 | 0.7924 | 0.7607 | 0.7621 | 0.6871 |
| 24 | 0.7982 | 0.5054 | 0.5054 | 0.5054 | 0.5228 | 0.7843 | 0.7643 | 0.7611 | 0.6854 |
| 25 | 0.7838 | 0.5054 | 0.5054 | 0.5054 | 0.5117 | 0.7945 | 0.7633 | 0.7772 | 0.6414 |
| 26 | 0.8302 | 0.5011 | 0.5011 | 0.5011 | 0.5165 | 0.8159 | 0.7352 | 0.6746 | 0.7355 |
| 27 | 0.7832 | 0.5026 | 0.5026 | 0.5026 | 0.5184 | 0.7939 | 0.7701 | 0.8032 | 0.6695 |
| 28 | 0.7165 | 0.5024 | 0.5024 | 0.5024 | 0.5134 | 0.7030 | 0.7491 | 0.6381 | 0.6305 |
| 29 | 0.7825 | 0.5028 | 0.5028 | 0.5028 | 0.5333 | 0.7445 | 0.7749 | 0.6567 | 0.6026 |
| 30 | 0.7650 | 0.5043 | 0.5043 | 0.5043 | 0.5179 | 0.7884 | 0.7517 | 0.7464 | 0.6623 |
| 31 | 0.6681 | 0.5058 | 0.5060 | 0.5060 | 0.5152 | 0.6410 | 0.6539 | 0.6639 | 0.6328 |
| 32 | 0.7232 | 0.5041 | 0.5042 | 0.5042 | 0.5004 | 0.6341 | 0.6755 | 0.6558 | 0.6327 |
| 33 | 0.7121 | 0.5040 | 0.5041 | 0.5041 | 0.5106 | 0.6415 | 0.6894 | 0.6805 | 0.6691 |
| 34 | 0.6840 | 0.5073 | 0.5073 | 0.5073 | 0.5085 | 0.6066 | 0.6661 | 0.6253 | 0.5930 |
| 35 | 0.7289 | 0.5027 | 0.5024 | 0.5024 | 0.5351 | 0.6695 | 0.6881 | 0.5908 | 0.6446 |
| 36 | 0.7413 | 0.5073 | 0.5072 | 0.5072 | 0.5219 | 0.6135 | 0.6318 | 0.6276 | 0.6802 |
| 37 | 0.7577 | 0.5018 | 0.5017 | 0.5017 | 0.5072 | 0.5412 | 0.5440 | 0.5282 | 0.5606 |
| 38 | 0.7414 | 0.5006 | 0.5006 | 0.5006 | 0.5106 | 0.5069 | 0.5018 | 0.5728 | 0.6174 |
| 39 | 0.6999 | 0.5030 | 0.5031 | 0.5031 | 0.5118 | 0.5621 | 0.6038 | 0.6693 | 0.6367 |
| 40 | 0.8505 | 0.5184 | 0.5184 | 0.5184 | 0.5122 | 0.6437 | 0.5796 | 0.8374 | 0.6690 |
| 41 | 0.7509 | 0.5146 | 0.5150 | 0.5150 | 0.5240 | 0.5235 | 0.5194 | 0.5450 | 0.7320 |
| 42 | 0.7528 | 0.5150 | 0.5150 | 0.5150 | 0.5266 | 0.5446 | 0.5613 | 0.7351 | 0.7171 |
| 43 | 0.8218 | 0.5151 | 0.5150 | 0.5150 | 0.5029 | 0.5128 | 0.5013 | 0.7132 | 0.7110 |
| 44 | 0.8392 | 0.5012 | 0.5011 | 0.5011 | 0.5102 | 0.6082 | 0.5245 | 0.7381 | 0.7201 |
| 45 | 0.8542 | 0.5019 | 0.5019 | 0.5019 | 0.5231 | 0.7425 | 0.5545 | 0.7801 | 0.6911 |
| 46 | 0.8633 | 0.5167 | 0.5168 | 0.5168 | 0.5102 | 0.7627 | 0.5658 | 0.7610 | 0.7356 |
| 47 | 0.8864 | 0.5049 | 0.5050 | 0.5050 | 0.5241 | 0.8041 | 0.5431 | 0.7905 | 0.7625 |
| 48 | 0.8777 | 0.5172 | 0.5175 | 0.5175 | 0.5214 | 0.8501 | 0.5913 | 0.7833 | 0.7563 |
| 49 | 0.8847 | 0.5110 | 0.5111 | 0.5111 | 0.5120 | 0.8651 | 0.5836 | 0.8214 | 0.7278 |
| 50 | 0.8942 | 0.5165 | 0.5168 | 0.5168 | 0.5358 | 0.8400 | 0.5553 | 0.7757 | 0.7390 |
| 51 | 0.8922 | 0.5128 | 0.5130 | 0.5130 | 0.5111 | 0.8397 | 0.6072 | 0.7992 | 0.7251 |
| 52 | 0.8990 | 0.5025 | 0.5025 | 0.5025 | 0.5180 | 0.8450 | 0.5930 | 0.7795 | 0.7243 |
| Avg. | 0.8029 | 0.5055 | 0.5056 | 0.5056 | 0.5203 | 0.7288 | 0.6658 | 0.7069 | 0.6910 |
| Std. | 0.0625 | 0.0053 | 0.0053 | 0.0053 | 0.0091 | 0.1072 | 0.0939 | 0.0876 | 0.0503 |

**Table 2:** Comparative AUC Analyses of Proposed and Benchmark Methods

For k-NN in Table 2, we followed the strategy by Hand et al. (2001) to set the value of k. Specifically, each possible k value was evaluated with 10-fold cross validation using training data; we selected the k value that yielded the best average AUC across the 10 folds. The SVM method in Table 2 used RBF kernel (Burges 1998) and was implemented with the standard software package LIBSVM (Chang and Lin 2011). We experimentally tuned the SVM parameters according to the commonly accepted guidelines (Hsu et al. 2003). We also benchmarked our method against SVM with other kernels including linear, polynomial, and sigmoid (Burges 1998). The Wilcoxon testing results suggest our method significantly outperforming SVM with any investigated kernel ($p < 0.001$). We do not include detailed AUCs of SVM with linear, polynomial, and sigmoid kernels because of space consideration.

Several observations warrant attention. First, the AUCs of the cascade methods and the influence probability method are barely greater than 0.5, which suggests their predictive power only marginally higher than that of random guess (Fawcett 2006). The poor performance of these methods partly reflects their exclusive reliance on social influence for adoption probability predictions (Chen et al. 2009, Goyal et al. 2010). Consequently, they predict that the adoption probability of a social entity in week $T + 1$ is zero if that entity has zero neighbors who adopt in week $T$, as in equation (47). Figure 7 shows the percentage of adopters (nonadopters) in week $T + 1$ who has zero neighbors adopting in week $T$. Across all 50 evaluations, on average 95.78% of adopters in week $T + 1$ has zero neighbors who adopt in week $T$ and 98.51% of nonadopters in week $T + 1$ has zero neighbors who adopt in week $T$. Therefore, on average, adoption probabilities of 95.78% of adopters and 98.51% of nonadopters are predicted as zero by these benchmark methods; and hence these social entities cannot be differentiated by them, which explains their performance. Low weekly adoption rates (0.42% on average) make these high percentages unsurprising. We note that adoption rates in our data are not uncommon; comparable or lower adoption rates have been reported by prior studies (Aral et al. 2009, Iyengar et al. 2011). Furthermore, similar percentages of adopters (nonadopters) having zero adopter neighbors have also been noted (Aral et al. 2009). The proposed method, on the other hand, considers more comprehensive set of factors underlying adoption decision; therefore, it can predict adoption probabilities more effectively.

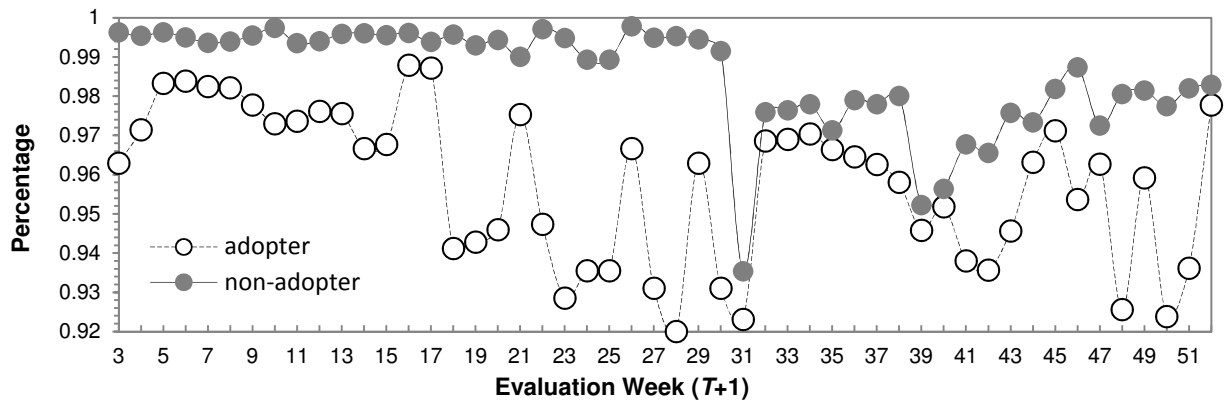

**Figure 7**: Percentage of Zero Neighbors Adopting in Week $T$

Second, the proposed method outperforms the benchmarked classification methods: NB, LWNB, SVM, and k-NN. The superiority of the proposed method over the benchmarked classification methods can be attributed to the consideration of hidden confounding power by the proposed method but not by the classification methods. Thus, experimental results in Table 2 also highlight the importance of hidden confounding power (and hence confounding factors) for predicting adoption probabilities. Both LWNB and the proposed method are NB-based classification methods. Methodologically, NB (Mitchell 1997) does not use local learning nor consider hidden confounding power; LWNB (Frank et al. 2003) employs local learning but does not consider hidden confounding power. Applying the Wilcoxon test to the AUCs of NB and LWNB, testing results suggest that NB significantly outperforms LWNB ($p < 0.01$). The performance of local learning methods depends on appropriate weights of training records (Atkeson et al. 1997). For our study, the hidden confounding power is unobserved, and LWNB weights training records using only observed (partial) variables, so training records may not be properly weighted, which could lead to the performance differential we observed. With local learning and consideration of hidden confounding power, the proposed method avoids this problem associated with LWNB and therefore can yield better performance than both NB and LWNB.

Third, as we report in Table 2, the ratio of the standard deviation to the mean for LEMNB is higher than that for cascade methods; however the ratio for LEMNB is comparable or lower than that for NB, LWNB, SVM, and k-NN. Note that LEMNB, NB, LWNB, SVM, and k-NN are learning-based predictive methods. These methods learn adoption patterns from training data and then leverage such patterns to predict future adoption behaviors. The deviation between prior adoption patterns and future adoption behaviors might affect the performance of these methods. If the deviation increases, the performance of a learning-based method is likely to decline; if the deviation decreases, its performance could improve. Thus, for LEMNB, NB, LWNB, SVM, and k-NN, we observe their performance varying across 50 evaluation weeks, though on average these methods (especially LEMNB) demonstrate good predictive performance, as suggested by their high average AUC scores. Cascade methods, on the other hand, predict the adoption probability of a social entity in week $T+1$ based on his or her neighbors who adopt in week $T$. As we have discussed, the high percentage of adopters (nonadopters) having zero neighbors adopting in week $T$ explains the poor performance of cascade methods. Thus, although the performance of cascade methods is relatively stable across evaluation weeks, they are stable at a very low performance level. In Appendix E, we plot the AUCs of each method across 50 evaluations.

Our evaluation results with the mobile social network show that the proposed method substantially outperforms all benchmark methods. Similar results emerge from our comparative evaluations with the avatar social network, as we detail in Appendix D. This superior performance derives from our consideration of a more comprehensive set of factors underlying adoption decision, including unobserved confounding factors. Our results also indicate that the predictive power of cascade methods that consider social influence solely is limited and that confounding factors are critical to effective adoption probability predictions.

## 5. Discussion and Conclusion

We take a data-driven approach to study adoption behaviors in a social network by predicting individuals' adoption probabilities from observed adoption data. From the lens of established social network theories, we identify and operationalize key factors underlying adoption decision, and then develop the locally-weighted EM method for Naïve Bayes learning to predict adoption probabilities on the basis of these key factors. Our study makes several research contributions. First, we develop a method to predict adoption probabilities by considering a more comprehensive set of factors underlying adoption decision than do existing methods. An essential novelty of our method is the consideration of unobserved confounding factors for predicting adoption probabilities. Second, we evaluate the proposed method with real-world data from two large-scale social networks and produce empirical evidence that reveals greater predictive power of the proposed method over all benchmark methods across the two social networks we studied. Third, our evaluation results shed light on the significance of confounding factors in adoption probability predictions; they further suggest that adoption probabilities should be predicted with factors beyond social influence. Our findings support and reinforce the motivation of the proposed method, i.e., better predicting adoption probabilities with a more comprehensive set of key factors underlying adoption decision, including confounding factors.

Our findings suggest that cascade methods relying on exclusive use of social influence seems limited in predictive power. This is intriguing because previous research shows the important effects of social influence (Pan et al. 2011, Altshuler et al. 2012, Pickard et al. 2011), while pointing out the significance of other forces. Forces above and beyond social influence are also recognized by (Bakshy et al. 2011, Watts and Peretti 2007), congruent with our approach. Overall, our results shed light on potential limitations of cascade methods solely using social influence rather than defying the value of social influence for predictions. Our findings also suggest the consideration of other forces, particularly unobserved confounding factors, which could augment the predictive power of methods emphasizing social influence. In particular, our study offers a viable way to operationalize confounding factors for predicting individuals' adoption probabilities in a network enabled by social media.

Our study also has several implications for practice. First, firms can use our method to enhance their social network-based target marketing efforts by better promoting product (service) adoptions in a network enabled by social media. Different from traditional target marketing, social network-based target marketing leverages essential structural linkage and interactions among individuals in a social network (Hill et al. 2006). Supported by our method, social commerce firms and online gaming providers alike can predict individual adoption probabilities on the basis of their social, demographical, and behavioral information, and select a subset of customers to focus on in each time period. By ranking potential adopters by their probabilities, a firm can differentiates consumers and designs personalized incentives, in light of each individual's adoption probability, rather than offering a uniform incentive. Firms could approach top-ranked potential adopters with strong cross- or up-selling strategies to generate more revenues and provide less likely potential adopters with incentives that lure them into adopting. Second, social commerce firms and online game providers also can use our method to estimate aggregate demand for their offerings over time. By summing adoption probabilities across potential adopters, a firm could predict the expected number of adopters in the next time period. Such prediction allows firms to gauge whether an offering is likely to go viral, which has crucial implications for their business decisions (Altshuler et al. 2012, Bandari et al. 2012). For example, if an offering is likely to go viral, the firm could act proactively to leverage the anticipated viral with appropriate marketing strategies, such as bundling or cross-selling. In addition, with effective estimates by our method, firms could allocate their resources and capabilities dynamically and intelligently across different time periods to improve performance and utilization efficiency. For example, firms can allocate more resources for time periods in which they predict a greater number of consumers will adopt their services. Furthermore, our method supports viral marketing, which requires identifying a target set of social entities whose adoptions will trigger the greatest spread of adoption throughout a social network. Identifying such seeds involves effective adoption probability predictions. Supported by our method, firms can become more effective in seed selection for viral marketing and perform the selection dynamically over time.

Our study could be extended in several directions. First, although the proposed method considers more comprehensive set of factors underlying individuals' adoption decisions than existing methods, there could be other factors not considered by the method, such as viral product features (Aral and Walker 2011), strength of weak ties (Granovetter 1973), and the connectedness of adopter neighbors (Backstrom et al. 2006). Future research should extend the proposed method by exploring and incorporating additional important factors and evaluate the effectiveness accordingly. In Appendix F, we provide a preliminary study that illustrates how to incorporate the connectedness of adopter neighbors into our method. Future research should also consider how to utilize useful information about hidden confounding power for better initialization of $\bar{\lambda}_{H|a}$, which in turn could improve the performance of the proposed method. Second, while our study provides a preliminary analysis of the performance differential between LWNB and NB, systematic methodological analyses as well as in-depth empirical evaluations are needed to analyze situations in which LWNB performs differently from NB. Third, prior studies have explored the interaction effects of social influence and entity similarity on adoption decisions (Aral et al. 2009, Aral and Walker 2012). In light of these studies, future research should examine how to extend the proposed method by incorporating such effects. Fourth, it is interesting to investigate alternative operationalizations of the factors underlying adoption decision and evaluate our method accordingly. A preliminary exploration in this direction is given in Appendix G. Finally, the current implementation of our method may not be scalable to social networks with millions of nodes, primarily due to the method's pairwise distance calculation, which requires $O(n^2)$ time under current implementation, where *n* is the number of nodes in a social network. Toward that end, use of a hierarchical tree-based distance calculation algorithm can reduce the time requirement for pairwise distance calculations from $O(n^2)$ to $O(n \log n)$ (Barnes and Hut 1986). The computational time requirements can be further reduced through parallel computing. Therefore, it is important to investigate how to reduce the time required by the proposed method for large social networks with millions of nodes. Finally, conducting a case study is worthy of future research attention, such as those in the form of social network-based target marketing.

**Appendix A: Illustrative Examples**

**A1. Euclidean distance of structural equivalence**

Figure A1 depicts an example social network at time *t*: the set of social entities $V = \{v_1, v_2, \cdots, v_6\}$ and nondirectional social ties connecting these social entities.

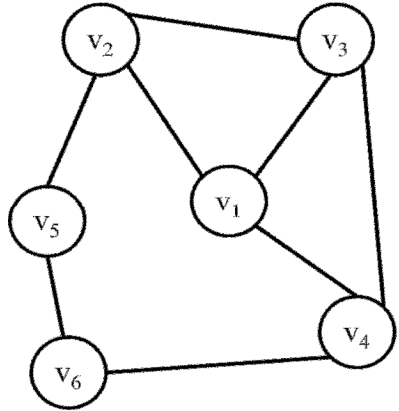

**Figure A1:** An Example Social Network

According to (4), the Euclidean distance of structural equivalence between $v_1$ and $v_5$ at time *t*, $y_{15}^t$ is calculated as the following,

$$y_{15}^t = \sqrt{(l_{12}^t - l_{52}^t)^2 + (l_{13}^t - l_{53}^t)^2 + (l_{14}^t - l_{54}^t)^2 + (l_{16}^t - l_{56}^t)^2}$$
$$= \sqrt{(1-1)^2 + (1-0)^2 + (1-0)^2 + (0-1)^2} = 1.7321.$$

**A2. Construction of training data from observed adoption information**

Considering the set of social entities $V = \{v_1, v_2, \cdots, v_6\}$ in a social network, we observe the following adoption information at current time $T = 3$: adopters $V_A^3 = \{v_1, v_2, v_3, v_4\}$, nonadopters $V_N^3 = \{v_5, v_6\}$, and adoption time for each adopter: $\tau_1 = 2, \tau_2 = 1, \tau_3 = 2, \tau_4 = 3$. According to adoption time, the earliest adopter is $v_2$, followed by $v_1$ and $v_3$ and then followed by $v_4$. Applying the algorithm in Figure 1, we show step-by-step results in Table A1.

Initially, training data are empty. The first for-loop computes powers from the earliest adopter to the latest adopter. Hence, powers on $v_2$ are computed first. Specifically, $I_2^{\tau_2-1} = I_2^0 = \sum_{v_i \in V_{early}} I_{i2}^0 = 0$ because $V_{early} = \phi$ (i.e., no earlier adopter before $v_2$). Similarly, $E_2^0 = 0$ and $S_2^0 = 0$. $A_2^1 = 1$ because $v_2$ adopts at time 1. Thus, $\{< I_2^0, E_2^0, S_2^0, A_2^1 >\} = \{< 0,0,0,1 >\}$ is added to training data. Social entity $v_2$ who adopts at time 1, termed as innovator by Bass (1969), receives no influence power, equivalence power, or similarity power; and the adoption of an innovator is driven by hidden confounding power (Van den Bulte and Lilien 2001).

Social entities $v_1$ and $v_3$ adopt at the same time and we random choose to compute powers on $v_1$ next. In particular, $I_1^{\tau_1-1} = I_1^1 = \sum_{v_i \in V_{early}} I_{i1}^1 = \sum_{v_i \in \{v_2\}} I_{i1}^1 = I_{21}^1$ because $v_2$ is the only earlier adopter before $v_1$. Similarly, we can calculate $E_1^1$ and $S_1^1$ according to their respective formula in the algorithm and add $<I_1^1, E_1^1, S_1^1, 1>$ to training data. Computing powers on $v_3$ and $v_4$ is similar to computing powers on $v_1$ and not discussed. The second for-loop calculates powers on each nonadopter. Taking $v_5$ as an example, $I_5^{T-1} = I_5^2 = \sum_{v_i \in V_{early}} I_{i5}^2 = \sum_{v_i \in \{v_1, v_3, v_2\}} I_{i5}^2 = I_{15}^2 + I_{25}^2 + I_{35}^2$. We then can calculate $E_5^2$ and $S_5^2$ according to their respective formula in the algorithm and add $<I_5^2, E_5^2, S_5^2, 0>$ to training data.

|  | *clock* | $V_{clk}$ | $V_{early}$ | *TRAIN* |
|---|---|---|---|---|
| Initial | 0 | $\phi$ | $\phi$ | $\phi$ |
| $v_2$ | 1 | $\{v_2\}$ | $\phi$ | $\{<0,0,0,1>\}$ |
| $v_1$ | 2 | $\{v_1\}$ | $\{v_2\}$ | $\{<0,0,0,1>, <I_1^1, E_1^1, S_1^1, 1>\}$ |
| $v_3$ | 2 | $\{v_1, v_3\}$ | $\{v_2\}$ | $\{<0,0,0,1>, <I_1^1, E_1^1, S_1^1, 1>, <I_3^1, E_3^1, S_3^1, 1>\}$ |
| $v_4$ | 3 | $\{v_4\}$ | $\{v_1, v_3, v_2\}$ | $\{<0,0,0,1>, <I_1^1, E_1^1, S_1^1, 1>, <I_3^1, E_3^1, S_3^1, 1>, <I_4^2, E_4^2, S_4^2, 1>\}$ |
| $v_5$ | 3 | $\{v_4\}$ | $\{v_1, v_3, v_2\}$ | $\{<0,0,0,1>, <I_1^1, E_1^1, S_1^1, 1>, <I_3^1, E_3^1, S_3^1, 1>, <I_4^2, E_4^2, S_4^2, 1>, <I_5^2, E_5^2, S_5^2, 0>\}$ |
| $v_6$ | 3 | $\{v_4\}$ | $\{v_1, v_3, v_2\}$ | $\{<0,0,0,1>, <I_1^1, E_1^1, S_1^1, 1>, <I_3^1, E_3^1, S_3^1, 1>, <I_4^2, E_4^2, S_4^2, 1>, <I_5^2, E_5^2, S_5^2, 0>, <I_6^2, E_6^2, S_6^2, 0>\}$ |

**Table A1:** An Example Illustrating the Construction of Training Data

**Appendix B: Derivation of (19)**

We need to set $K$ such that for all $i = 1, 2, \cdots, n$,

$$\iint \left[K - C_i - (H_i - H_q)^2\right] f(H_i|A_i, \overline{\boldsymbol{\theta}}) f(H_q|\overline{\boldsymbol{\theta}}) dH_i dH_q > 0.$$

That is, for all $i = 1, 2, \cdots, n$,

$$K > C_i + \iint (H_i - H_q)^2 f(H_i|A_i, \overline{\boldsymbol{\theta}}) f(H_q|\overline{\boldsymbol{\theta}}) dH_i dH_q.$$

Given $\overline{\boldsymbol{\theta}}$, densities $f(H_i|A_i, \overline{\boldsymbol{\theta}})$ and $f(H_q|\overline{\boldsymbol{\theta}})$ are known. For example, probability density of $H_i$ given $A_i = 1$ is $f(H_i|A_i = 1, \overline{\boldsymbol{\theta}}) = \bar{\lambda}_{H|1} \exp(-\bar{\lambda}_{H|1} H_i)$.

Applying $f(H_q|\overline{\boldsymbol{\theta}}) = f(H_q|A_q = 1, \overline{\boldsymbol{\theta}}) P(A_q = 1|\overline{\boldsymbol{\theta}}) + f(H_q|A_q = 0, \overline{\boldsymbol{\theta}}) P(A_q = 0|\overline{\boldsymbol{\theta}})$ and solving the integral, we have, for all $i = 1, 2, \cdots, n$,

$$K > C_i + \frac{2}{\bar{\lambda}_{H|A_i}^2} - \frac{2\bar{p}_1}{\bar{\lambda}_{H|A_i}\bar{\lambda}_{H|1}} + \frac{2\bar{p}_1}{\bar{\lambda}_{H|1}^2} - \frac{2\bar{p}_0}{\bar{\lambda}_{H|A_i}\bar{\lambda}_{H|0}} + \frac{2\bar{p}_0}{\bar{\lambda}_{H|0}^2}.$$

Considering that $A_i$ can only be 0 or 1, we need to set $K$ such that for all $i = 1, 2, \cdots, n$,

$$K > C_i + \frac{2}{\bar{\lambda}_{H|0}^2} + \frac{2\bar{p}_1}{\bar{\lambda}_{H|1}^2} - \frac{2\bar{p}_1}{\bar{\lambda}_{H|0}\bar{\lambda}_{H|1}} \quad \text{if} \quad A_i = 0,$$

or

$$K > C_i + \frac{2}{\bar{\lambda}_{H|1}^2} + \frac{2\bar{p}_0}{\bar{\lambda}_{H|0}^2} - \frac{2\bar{p}_0}{\bar{\lambda}_{H|0}\bar{\lambda}_{H|1}} \quad \text{if} \quad A_i = 1.$$

It is easy to show that the above constraint is satisfied if we set $K$ as,

$$K = C_i^{max} + \frac{2}{\bar{\lambda}_{H|0}^2} + \frac{2}{\bar{\lambda}_{H|1}^2},$$

where $C_i^{max}$ is the maximum among all $C_i$, $i = 1, 2, \cdots, n$.

## Appendix C: Proofs of Theorems

### C1. Proof of Theorem 1

Let $g(\boldsymbol{\theta}) = E_{H_i, H_q | \bar{\boldsymbol{\theta}}} \{\sum_{i=1}^n W_i \ln[P(D_i|\boldsymbol{\theta})]\}$. According to (16), we have

$$g(\boldsymbol{\theta}) = \sum_{i=1}^n \iint \left[K - C_i - (H_i - H_q)^2\right] \ln[P(D_i|\boldsymbol{\theta})] f(H_i|A_i, \bar{\boldsymbol{\theta}}) f(H_q|\bar{\boldsymbol{\theta}}) dH_i dH_q$$

$$= \sum_{i=1}^n \iint \left[K - C_i - (H_i - H_q)^2\right] \ln[P(I_i, E_i, S_i, H_i, A_i|\boldsymbol{\theta})] f(H_i|A_i, \bar{\boldsymbol{\theta}}) f(H_q|\bar{\boldsymbol{\theta}}) dH_i dH_q$$

$$= \sum_{i=1}^n \iint \left[K - C_i - (H_i - H_q)^2\right] \{\ln[P(I_i|A_i, \boldsymbol{\theta})] + \ln[P(E_i|A_i, \boldsymbol{\theta})] + \ln[P(S_i|A_i, \boldsymbol{\theta})]$$

$$+ \ln[P(H_i|A_i, \boldsymbol{\theta})] + \ln[P(A_i|\boldsymbol{\theta})]\} f(H_i|A_i, \bar{\boldsymbol{\theta}}) f(H_q|\bar{\boldsymbol{\theta}}) dH_i dH_q$$

(by conditional independence of Naïve Bayes)

$$= \underbrace{\sum_{i=1}^n \int\int \left[K - C_i - (H_i - H_q)^2\right] \ln P(I_i|A_i, \boldsymbol{\theta}) f(H_i|A_i, \bar{\boldsymbol{\theta}}) f(H_q|\bar{\boldsymbol{\theta}}) dH_i dH_q}_{F1} +$$

$$\underbrace{\sum_{i=1}^n \int\int \left[K - C_i - (H_i - H_q)^2\right] \ln P(E_i|A_i, \boldsymbol{\theta}) f(H_i|A_i, \bar{\boldsymbol{\theta}}) f(H_q|\bar{\boldsymbol{\theta}}) dH_i dH_q}_{F2} +$$

$$\underbrace{\sum_{i=1}^n \int\int \left[K - C_i - (H_i - H_q)^2\right] \ln P(S_i|A_i, \boldsymbol{\theta}) f(H_i|A_i, \bar{\boldsymbol{\theta}}) f(H_q|\bar{\boldsymbol{\theta}}) dH_i dH_q}_{F3} +$$

$$\underbrace{\sum_{i=1}^n \int\int \left[K - C_i - (H_i - H_q)^2\right] \ln P(H_i|A_i, \boldsymbol{\theta}) f(H_i|A_i, \bar{\boldsymbol{\theta}}) f(H_q|\bar{\boldsymbol{\theta}}) dH_i dH_q}_{F4} +$$

$$\underbrace{\sum_{i=1}^n \int\int \left[K - C_i - (H_i - H_q)^2\right] \ln P(A_i|\boldsymbol{\theta}) f(H_i|A_i, \bar{\boldsymbol{\theta}}) f(H_q|\bar{\boldsymbol{\theta}}) dH_i dH_q}_{F5}.$$

The integrals $F1$, $F2$, $F3$, $F4$, and $F5$ can be further expressed as below.

$$F1 = \sum_{i=1}^n \left\{ A_i \left[ (\ln \lambda_{I|1} - \lambda_{I|1} I_i) \left( K - C_i - \frac{2}{\bar{\lambda}_{H|1}^2} - 2\bar{p}_0 \left( \frac{1}{\bar{\lambda}_{H|0}^2} - \frac{1}{\bar{\lambda}_{H|1} \bar{\lambda}_{H|0}} \right) \right) \right] \right.$$

$$\left. + (1 - A_i) \left[ (\ln \lambda_{I|0} - \lambda_{I|0} I_i) \left( K - C_i - \frac{2}{\bar{\lambda}_{H|0}^2} - 2\bar{p}_1 \left( \frac{1}{\bar{\lambda}_{H|1}^2} - \frac{1}{\bar{\lambda}_{H|1} \bar{\lambda}_{H|0}} \right) \right) \right] \right\}$$

$$F2 = \sum_{i=1}^{n} \left\{ A_i \left[ (\ln \lambda_{E|1} - \lambda_{E|1} E_i) \left( K - C_i - \frac{2}{\bar{\lambda}_{H|1}^2} - 2\bar{p}_0 \left( \frac{1}{\bar{\lambda}_{H|0}^2} - \frac{1}{\bar{\lambda}_{H|1}\bar{\lambda}_{H|0}} \right) \right) \right] \right.$$

$$\left. + (1 - A_i) \left[ (\ln \lambda_{E|0} - \lambda_{E|0} E_i) \left( K - C_i - \frac{2}{\bar{\lambda}_{H|0}^2} - 2\bar{p}_1 \left( \frac{1}{\bar{\lambda}_{H|1}^2} - \frac{1}{\bar{\lambda}_{H|1}\bar{\lambda}_{H|0}} \right) \right) \right] \right\}$$

$$F3 = \sum_{i=1}^{n} \left\{ A_i \left[ (\ln \lambda_{S|1} - \lambda_{S|1} S_i) \left( K - C_i - \frac{2}{\bar{\lambda}_{H|1}^2} - 2\bar{p}_0 \left( \frac{1}{\bar{\lambda}_{H|0}^2} - \frac{1}{\bar{\lambda}_{H|1}\bar{\lambda}_{H|0}} \right) \right) \right] \right.$$

$$\left. + (1 - A_i) \left[ (\ln \lambda_{S|0} - \lambda_{S|0} S_i) \left( K - C_i - \frac{2}{\bar{\lambda}_{H|0}^2} - 2\bar{p}_1 \left( \frac{1}{\bar{\lambda}_{H|1}^2} - \frac{1}{\bar{\lambda}_{H|1}\bar{\lambda}_{H|0}} \right) \right) \right] \right\}$$

$$F4 = \sum_{i=1}^{n} \left\{ A_i \left[ (K - C_i) \ln \lambda_{H|1} - \frac{2}{\bar{\lambda}_{H|1}^2} \ln \lambda_{H|1} - 2\bar{p}_0 \left( \frac{1}{\bar{\lambda}_{H|0}^2} - \frac{1}{\bar{\lambda}_{H|1}\bar{\lambda}_{H|0}} \right) \ln \lambda_{H|1} - \frac{(K - C_i)}{\bar{\lambda}_{H|1}} \lambda_{H|1} \right. \right.$$

$$\left. - \frac{2\bar{p}_1}{\bar{\lambda}_{H|1}\bar{\lambda}_{H|0}^2} \lambda_{H|1} + \frac{4\bar{p}_1}{\bar{\lambda}_{H|0}\bar{\lambda}_{H|1}^2} \lambda_{H|1} + \frac{6 - 2\bar{p}_1}{\bar{\lambda}_{H|1}^3} \lambda_{H|1} \right]$$

$$+ (1 - A_i) \left[ (K - C_i) \ln \lambda_{H|0} - \frac{2}{\bar{\lambda}_{H|0}^2} \ln \lambda_{H|0} - 2\bar{p}_1 \left( \frac{1}{\bar{\lambda}_{H|1}^2} - \frac{1}{\bar{\lambda}_{H|1}\bar{\lambda}_{H|0}} \right) \ln \lambda_{H|0} - \frac{(K - C_i)}{\bar{\lambda}_{H|0}} \lambda_{H|0} \right.$$

$$\left. \left. - \frac{2\bar{p}_0}{\bar{\lambda}_{H|0}\bar{\lambda}_{H|1}^2} \lambda_{H|0} + \frac{4\bar{p}_0}{\bar{\lambda}_{H|1}\bar{\lambda}_{H|0}^2} \lambda_{H|0} + \frac{6 - 2\bar{p}_0}{\bar{\lambda}_{H|0}^3} \lambda_{H|0} \right] \right\}$$

$$F5 = \sum_{i=1}^{n} \left[ A_i \ln p_1 \left( K - C_i - \frac{2}{\bar{\lambda}_{H|1}^2} - 2\bar{p}_0 \left( \frac{1}{\bar{\lambda}_{H|0}^2} - \frac{1}{\bar{\lambda}_{H|1}\bar{\lambda}_{H|0}} \right) \right) \right.$$

$$\left. + (1 - A_i) \ln p_0 \left( K - C_i - \frac{2}{\bar{\lambda}_{H|0}^2} - 2\bar{p}_1 \left( \frac{1}{\bar{\lambda}_{H|1}^2} - \frac{1}{\bar{\lambda}_{H|1}\bar{\lambda}_{H|0}} \right) \right) \right]$$

Recalling that $\boldsymbol{\theta} = <p_1, \lambda_{I|1}, \lambda_{E|1}, \lambda_{S|1}, \lambda_{H|1}, p_0, \lambda_{I|0}, \lambda_{E|0}, \lambda_{S|0}, \lambda_{H|0}>$, the Hessian matrix of $g(\boldsymbol{\theta})$ is given by (Greene 2008),

$$\begin{bmatrix} \frac{\partial^2 g(\boldsymbol{\theta})}{\partial p_1 \, \partial p_1} & \frac{\partial^2 g(\boldsymbol{\theta})}{\partial p_1 \, \partial \lambda_{I|1}} & \cdots & \frac{\partial^2 g(\boldsymbol{\theta})}{\partial p_1 \, \partial \lambda_{H|0}} \\ \frac{\partial^2 g(\boldsymbol{\theta})}{\partial \lambda_{I|1} \, \partial p_1} & \frac{\partial^2 g(\boldsymbol{\theta})}{\partial \lambda_{I|1} \, \partial \lambda_{I|1}} & \cdots & \frac{\partial^2 g(\boldsymbol{\theta})}{\partial \lambda_{I|1} \, \partial \lambda_{H|0}} \\ \vdots & \vdots & \vdots & \vdots \\ \frac{\partial^2 g(\boldsymbol{\theta})}{\partial \lambda_{H|0} \, \partial p_1} & \frac{\partial^2 g(\boldsymbol{\theta})}{\partial \lambda_{H|0} \, \partial \lambda_{I|1}} & \cdots & \frac{\partial^2 g(\boldsymbol{\theta})}{\partial \lambda_{H|0} \, \partial \lambda_{H|0}} \end{bmatrix}$$

___

We first calculate first-order partial derivatives of $g(\boldsymbol{\theta})$ w.r.t. each parameter in $\boldsymbol{\theta}$.

$$\frac{\partial g(\boldsymbol{\theta})}{\partial p_1} = \sum_{i=1}^{n} \left\{ \frac{A_i}{p_1} \left[ K - C_i - \frac{2}{\bar{\lambda}_{H|1}^2} - 2\bar{p}_0 \left( \frac{1}{\bar{\lambda}_{H|0}^2} - \frac{1}{\bar{\lambda}_{H|1}\bar{\lambda}_{H|0}} \right) \right] \right.$$
$$\left. - \frac{(1-A_i)}{1-p_1} \left[ K - C_i - \frac{2}{\bar{\lambda}_{H|0}^2} - 2\bar{p}_1 \left( \frac{1}{\bar{\lambda}_{H|1}^2} - \frac{1}{\bar{\lambda}_{H|1}\bar{\lambda}_{H|0}} \right) \right] \right\}$$

$$\frac{\partial g(\boldsymbol{\theta})}{\partial \lambda_{I|1}} = \sum_{i=1}^{n} A_i \left( \frac{1}{\lambda_{I|1}} - I_i \right) [K - C_i - \frac{2}{\bar{\lambda}_{H|1}^2} - 2\bar{p}_0 (\frac{1}{\bar{\lambda}_{H|0}^2} - \frac{1}{\bar{\lambda}_{H|1}\bar{\lambda}_{H|0}})]$$

$$\frac{\partial g(\boldsymbol{\theta})}{\partial \lambda_{E|1}} = \sum_{i=1}^{n} A_i \left( \frac{1}{\lambda_{E|1}} - E_i \right) [K - C_i - \frac{2}{\bar{\lambda}_{H|1}^2} - 2\bar{p}_0 (\frac{1}{\bar{\lambda}_{H|0}^2} - \frac{1}{\bar{\lambda}_{H|1}\bar{\lambda}_{H|0}})]$$

$$\frac{\partial g(\boldsymbol{\theta})}{\partial \lambda_{S|1}} = \sum_{i=1}^{n} A_i \left( \frac{1}{\lambda_{S|1}} - S_i \right) [K - C_i - \frac{2}{\bar{\lambda}_{H|1}^2} - 2\bar{p}_0 (\frac{1}{\bar{\lambda}_{H|0}^2} - \frac{1}{\bar{\lambda}_{H|1}\bar{\lambda}_{H|0}})]$$

$$\frac{\partial g(\boldsymbol{\theta})}{\partial \lambda_{H|1}} = \sum_{i=1}^{n} A_i \{ \frac{1}{\lambda_{H|1}} [K - C_i - \frac{2}{\bar{\lambda}_{H|1}^2} - 2\bar{p}_0 \left( \frac{1}{\bar{\lambda}_{H|0}^2} - \frac{1}{\bar{\lambda}_{H|1}\bar{\lambda}_{H|0}} \right)] - \frac{K - C_i}{\bar{\lambda}_{H|1}} - \frac{2\bar{p}_1}{\bar{\lambda}_{H|0}^2 \bar{\lambda}_{H|1}} + \frac{4\bar{p}_1}{\bar{\lambda}_{H|1}^2 \bar{\lambda}_{H|0}}$$
$$+ \frac{6 - 2\bar{p}_1}{\bar{\lambda}_{H|1}^3} \}$$

$$\frac{\partial g(\boldsymbol{\theta})}{\partial p_0} = \sum_{i=1}^{n} \left\{ \frac{-A_i}{1-p_0} \left[ K - C_i - \frac{2}{\bar{\lambda}_{H|1}^2} - 2\bar{p}_0 \left( \frac{1}{\bar{\lambda}_{H|0}^2} - \frac{1}{\bar{\lambda}_{H|1}\bar{\lambda}_{H|0}} \right) \right] \right.$$
$$\left. + \frac{(1-A_i)}{p_0} \left[ K - C_i - \frac{2}{\bar{\lambda}_{H|0}^2} - 2\bar{p}_1 \left( \frac{1}{\bar{\lambda}_{H|1}^2} - \frac{1}{\bar{\lambda}_{H|1}\bar{\lambda}_{H|0}} \right) \right] \right\}$$

$$\frac{\partial g(\boldsymbol{\theta})}{\partial \lambda_{I|0}} = \sum_{i=1}^{n} (1 - A_i) \left( \frac{1}{\lambda_{I|0}} - I_i \right) [K - C_i - \frac{2}{\bar{\lambda}_{H|0}^2} - 2\bar{p}_1 (\frac{1}{\bar{\lambda}_{H|1}^2} - \frac{1}{\bar{\lambda}_{H|1}\bar{\lambda}_{H|0}})]$$

$$\frac{\partial g(\boldsymbol{\theta})}{\partial \lambda_{E|0}} = \sum_{i=1}^{n} (1 - A_i) \left( \frac{1}{\lambda_{E|0}} - E_i \right) [K - C_i - \frac{2}{\bar{\lambda}_{H|0}^2} - 2\bar{p}_1 (\frac{1}{\bar{\lambda}_{H|1}^2} - \frac{1}{\bar{\lambda}_{H|1}\bar{\lambda}_{H|0}})]$$

$$\frac{\partial g(\boldsymbol{\theta})}{\partial \lambda_{S|0}} = \sum_{i=1}^{n} (1 - A_i) \left( \frac{1}{\lambda_{S|0}} - S_i \right) [K - C_i - \frac{2}{\bar{\lambda}_{H|0}^2} - 2\bar{p}_1 (\frac{1}{\bar{\lambda}_{H|1}^2} - \frac{1}{\bar{\lambda}_{H|1}\bar{\lambda}_{H|0}})]$$

$$\frac{\partial g(\boldsymbol{\theta})}{\partial \lambda_{H|0}} = \sum_{i=1}^{n} (1 - A_i) \{ \frac{1}{\lambda_{H|0}} [K - C_i - \frac{2}{\bar{\lambda}_{H|0}^2} - 2\bar{p}_1 \left( \frac{1}{\bar{\lambda}_{H|1}^2} - \frac{1}{\bar{\lambda}_{H|1}\bar{\lambda}_{H|0}} \right)] - \frac{K - C_i}{\bar{\lambda}_{H|0}} - \frac{2\bar{p}_0}{\bar{\lambda}_{H|1}^2 \bar{\lambda}_{H|0}} + \frac{4\bar{p}_0}{\bar{\lambda}_{H|0}^2 \bar{\lambda}_{H|1}}$$
$$+ \frac{6 - 2\bar{p}_0}{\bar{\lambda}_{H|0}^3} \}$$

___

For any first-order partial derivative on a parameter $x \in \boldsymbol{\theta}$, its second-order partial derivation on a different parameter $y \in \boldsymbol{\theta}, y \neq x$, yields 0 because each first-order partial derivative shown above involves no other parameter. That is, for $x, y \in \boldsymbol{\theta}$

$$\frac{\partial g(\boldsymbol{\theta})}{\partial x \, \partial y} = 0 \qquad \text{if } x \neq y.$$

Hence, the off-diagonal terms in the Hessian matrix are all zero. The main diagonal terms of the Hessian matrix are given below.

$$\frac{\partial^2 g(\boldsymbol{\theta})}{\partial p_1 \, \partial p_1} = \sum_{i=1}^{n} \{\frac{-A_i}{p_1^2} \left[ K - C_i - \frac{2}{\bar{\lambda}_{H|1}^2} - 2\bar{p}_0 \left( \frac{1}{\bar{\lambda}_{H|0}^2} - \frac{1}{\bar{\lambda}_{H|1} \bar{\lambda}_{H|0}} \right) \right]$$

$$- \frac{(1 - A_i)}{(1 - p_1)^2} \left[ K - C_i - \frac{2}{\bar{\lambda}_{H|0}^2} - 2\bar{p}_1 \left( \frac{1}{\bar{\lambda}_{H|1}^2} - \frac{1}{\bar{\lambda}_{H|1} \bar{\lambda}_{H|0}} \right) \right]\}$$

$$\frac{\partial^2 g(\boldsymbol{\theta})}{\partial \lambda_{I|1} \, \partial \lambda_{I|1}} = \sum_{i=1}^{n} \left( \frac{-A_i}{\lambda_{I|1}^2} \right) [K - C_i - \frac{2}{\bar{\lambda}_{H|1}^2} - 2\bar{p}_0 (\frac{1}{\bar{\lambda}_{H|0}^2} - \frac{1}{\bar{\lambda}_{H|1} \bar{\lambda}_{H|0}})]$$

$$\frac{\partial^2 g(\boldsymbol{\theta})}{\partial \lambda_{E|1} \, \partial \lambda_{E|1}} = \sum_{i=1}^{n} \left( \frac{-A_i}{\lambda_{E|1}^2} \right) [K - C_i - \frac{2}{\bar{\lambda}_{H|1}^2} - 2\bar{p}_0 (\frac{1}{\bar{\lambda}_{H|0}^2} - \frac{1}{\bar{\lambda}_{H|1} \bar{\lambda}_{H|0}})]$$

$$\frac{\partial^2 g(\boldsymbol{\theta})}{\partial \lambda_{S|1} \, \partial \lambda_{S|1}} = \sum_{i=1}^{n} \left( \frac{-A_i}{\lambda_{S|1}^2} \right) [K - C_i - \frac{2}{\bar{\lambda}_{H|1}^2} - 2\bar{p}_0 (\frac{1}{\bar{\lambda}_{H|0}^2} - \frac{1}{\bar{\lambda}_{H|1} \bar{\lambda}_{H|0}})]$$

$$\frac{\partial^2 g(\boldsymbol{\theta})}{\partial \lambda_{H|1} \, \partial \lambda_{H|1}} = \sum_{i=1}^{n} (\frac{-A_i}{\lambda_{H|1}^2})[K - C_i - \frac{2}{\bar{\lambda}_{H|1}^2} - 2\bar{p}_0 \left( \frac{1}{\bar{\lambda}_{H|0}^2} - \frac{1}{\bar{\lambda}_{H|1} \bar{\lambda}_{H|0}} \right)]$$

$$\frac{\partial^2 g(\boldsymbol{\theta})}{\partial p_0 \, \partial p_0} = \sum_{i=1}^{n} \{\frac{-(1 - A_i)}{p_0^2} \left[ K - C_i - \frac{2}{\bar{\lambda}_{H|0}^2} - 2\bar{p}_1 \left( \frac{1}{\bar{\lambda}_{H|1}^2} - \frac{1}{\bar{\lambda}_{H|1} \bar{\lambda}_{H|0}} \right) \right]$$

$$- \frac{A_i}{(1 - p_0)^2} \left[ K - C_i - \frac{2}{\bar{\lambda}_{H|1}^2} - 2\bar{p}_0 \left( \frac{1}{\bar{\lambda}_{H|0}^2} - \frac{1}{\bar{\lambda}_{H|1} \bar{\lambda}_{H|0}} \right) \right]\}$$

$$\frac{\partial^2 g(\boldsymbol{\theta})}{\partial \lambda_{I|0} \, \partial \lambda_{I|0}} = \sum_{i=1}^{n} \frac{-(1 - A_i)}{\lambda_{I|0}^2} [K - C_i - \frac{2}{\bar{\lambda}_{H|0}^2} - 2\bar{p}_1 (\frac{1}{\bar{\lambda}_{H|1}^2} - \frac{1}{\bar{\lambda}_{H|1} \bar{\lambda}_{H|0}})]$$

$$\frac{\partial^2 g(\boldsymbol{\theta})}{\partial \lambda_{E|0} \, \partial \lambda_{E|0}} = \sum_{i=1}^{n} \frac{-(1 - A_i)}{\lambda_{E|0}^2} [K - C_i - \frac{2}{\bar{\lambda}_{H|0}^2} - 2\bar{p}_1 (\frac{1}{\bar{\lambda}_{H|1}^2} - \frac{1}{\bar{\lambda}_{H|1} \bar{\lambda}_{H|0}})]$$

$$\frac{\partial^2 g(\boldsymbol{\theta})}{\partial \lambda_{S|0} \, \partial \lambda_{S|0}} = \sum_{i=1}^{n} \frac{-(1 - A_i)}{\lambda_{S|0}^2} [K - C_i - \frac{2}{\bar{\lambda}_{H|0}^2} - 2\bar{p}_1 (\frac{1}{\bar{\lambda}_{H|1}^2} - \frac{1}{\bar{\lambda}_{H|1} \bar{\lambda}_{H|0}})]$$

$$\frac{\partial^2 g(\boldsymbol{\theta})}{\partial \lambda_{H|0} \, \partial \lambda_{H|0}} = \sum_{i=1}^{n} \frac{-(1 - A_i)}{\lambda_{H|0}^2} [K - C_i - \frac{2}{\bar{\lambda}_{H|0}^2} - 2\bar{p}_1 (\frac{1}{\bar{\lambda}_{H|1}^2} - \frac{1}{\bar{\lambda}_{H|1} \bar{\lambda}_{H|0}})]$$

___

By (19), both $K - C_i - \frac{2}{\bar{\lambda}_{H|1}^2} - 2\bar{p}_0 \left(\frac{1}{\bar{\lambda}_{H|0}^2} - \frac{1}{\bar{\lambda}_{H|1}\bar{\lambda}_{H|0}}\right)$ and $K - C_i - \frac{2}{\bar{\lambda}_{H|0}^2} - 2\bar{p}_1 \left(\frac{1}{\bar{\lambda}_{H|1}^2} - \frac{1}{\bar{\lambda}_{H|1}\bar{\lambda}_{H|0}}\right)$ are positive. Hence, the above second-order partial derivatives are all negative. Therefore, the $k^{th}$ principal minor of the Hessian matrix is negative when $k$ is odd and positive when $k$ is even; and thus the Hessian matrix is negative definite (Dowling 1980). This completes the proof.

## C2. Proof of Theorem 2

Let $g(\boldsymbol{\theta}) = E_{H_i,H_q|\bar{\boldsymbol{\theta}}}\{\sum_{i=1}^n W_i \ln[P(D_i|\boldsymbol{\theta})]\}$. Taking first-order partial differentiation of $g(\boldsymbol{\theta})$ w.r.t. each parameter in $\boldsymbol{\theta} =< p_1, \lambda_{I|1}, \lambda_{E|1}, \lambda_{S|1}, \lambda_{H|1}, p_0, \lambda_{I|0}, \lambda_{E|0}, \lambda_{S|0}, \lambda_{H|0} >$, and making the obtained forms equal zero, we have,

$$\sum_{i=1}^n \{\frac{A_i}{p_1}\left[K - C_i - \frac{2}{\bar{\lambda}_{H|1}^2} - 2\bar{p}_0 \left(\frac{1}{\bar{\lambda}_{H|0}^2} - \frac{1}{\bar{\lambda}_{H|1}\bar{\lambda}_{H|0}}\right)\right] - \frac{(1-A_i)}{1-p_1}\left[K - C_i - \frac{2}{\bar{\lambda}_{H|0}^2} - 2\bar{p}_1 \left(\frac{1}{\bar{\lambda}_{H|1}^2} - \frac{1}{\bar{\lambda}_{H|1}\bar{\lambda}_{H|0}}\right)\right]\} = 0$$

$$\sum_{i=1}^n A_i \left(\frac{1}{\lambda_{I|1}} - I_i\right)[K - C_i - \frac{2}{\bar{\lambda}_{H|1}^2} - 2\bar{p}_0(\frac{1}{\bar{\lambda}_{H|0}^2} - \frac{1}{\bar{\lambda}_{H|1}\bar{\lambda}_{H|0}})] = 0$$

$$\sum_{i=1}^n A_i \left(\frac{1}{\lambda_{E|1}} - E_i\right)[K - C_i - \frac{2}{\bar{\lambda}_{H|1}^2} - 2\bar{p}_0(\frac{1}{\bar{\lambda}_{H|0}^2} - \frac{1}{\bar{\lambda}_{H|1}\bar{\lambda}_{H|0}})] = 0$$

$$\sum_{i=1}^n A_i \left(\frac{1}{\lambda_{S|1}} - S_i\right)[K - C_i - \frac{2}{\bar{\lambda}_{H|1}^2} - 2\bar{p}_0(\frac{1}{\bar{\lambda}_{H|0}^2} - \frac{1}{\bar{\lambda}_{H|1}\bar{\lambda}_{H|0}})] = 0$$

$$\sum_{i=1}^n A_i\{\frac{1}{\lambda_{H|1}}[K - C_i - \frac{2}{\bar{\lambda}_{H|1}^2} - 2\bar{p}_0 \left(\frac{1}{\bar{\lambda}_{H|0}^2} - \frac{1}{\bar{\lambda}_{H|1}\bar{\lambda}_{H|0}}\right)] - \frac{K-C_i}{\bar{\lambda}_{H|1}} - \frac{2\bar{p}_1}{\bar{\lambda}_{H|0}^2\bar{\lambda}_{H|1}} + \frac{4\bar{p}_1}{\bar{\lambda}_{H|1}^2\bar{\lambda}_{H|0}} + \frac{6-2\bar{p}_1}{\bar{\lambda}_{H|1}^3}\} = 0$$

$$\sum_{i=1}^n \{\frac{-A_i}{1-p_0}\left[K - C_i - \frac{2}{\bar{\lambda}_{H|1}^2} - 2\bar{p}_0 \left(\frac{1}{\bar{\lambda}_{H|0}^2} - \frac{1}{\bar{\lambda}_{H|1}\bar{\lambda}_{H|0}}\right)\right] + \frac{(1-A_i)}{1-p_0}\left[K - C_i - \frac{2}{\bar{\lambda}_{H|0}^2} - 2\bar{p}_1 \left(\frac{1}{\bar{\lambda}_{H|1}^2} - \frac{1}{\bar{\lambda}_{H|1}\bar{\lambda}_{H|0}}\right)\right]\} = 0$$

$$\sum_{i=1}^n (1 - A_i) \left(\frac{1}{\lambda_{I|0}} - I_i\right)[K - C_i - \frac{2}{\bar{\lambda}_{H|0}^2} - 2\bar{p}_1(\frac{1}{\bar{\lambda}_{H|1}^2} - \frac{1}{\bar{\lambda}_{H|1}\bar{\lambda}_{H|0}})] = 0$$

$$\sum_{i=1}^n (1 - A_i) \left(\frac{1}{\lambda_{E|0}} - E_i\right)[K - C_i - \frac{2}{\bar{\lambda}_{H|0}^2} - 2\bar{p}_1(\frac{1}{\bar{\lambda}_{H|1}^2} - \frac{1}{\bar{\lambda}_{H|1}\bar{\lambda}_{H|0}})] = 0$$

$$\sum_{i=1}^n (1 - A_i) \left(\frac{1}{\lambda_{S|0}} - S_i\right)[K - C_i - \frac{2}{\bar{\lambda}_{H|0}^2} - 2\bar{p}_1(\frac{1}{\bar{\lambda}_{H|1}^2} - \frac{1}{\bar{\lambda}_{H|1}\bar{\lambda}_{H|0}})] = 0$$

$$\sum_{i=1}^n (1-A_i)\{\frac{1}{\lambda_{H|0}}[K - C_i - \frac{2}{\bar{\lambda}_{H|0}^2} - 2\bar{p}_1 \left(\frac{1}{\bar{\lambda}_{H|1}^2} - \frac{1}{\bar{\lambda}_{H|1}\bar{\lambda}_{H|0}}\right)] - \frac{K-C_i}{\bar{\lambda}_{H|0}} - \frac{2\bar{p}_0}{\bar{\lambda}_{H|1}^2\bar{\lambda}_{H|0}} + \frac{4\bar{p}_0}{\bar{\lambda}_{H|0}^2\bar{\lambda}_{H|1}} + \frac{6-2\bar{p}_0}{\bar{\lambda}_{H|0}^3}\} = 0$$

We note that the first-order partial derivatives in the above equations are obtained during the proof of Theorem 1.

Let

$$Q_i = (K - C_i)\bar{\lambda}^2_{H|0}\bar{\lambda}^2_{H|1} - 2\bar{\lambda}^2_{H|0} - 2\bar{p}_0(\bar{\lambda}^2_{H|1} - \bar{\lambda}_{H|0}\bar{\lambda}_{H|1}),$$

$$R_i = (K - C_i)\bar{\lambda}^2_{H|0}\bar{\lambda}^2_{H|1} - 2\bar{\lambda}^2_{H|1} - 2\bar{p}_1(\bar{\lambda}^2_{H|0} - \bar{\lambda}_{H|0}\bar{\lambda}_{H|1}).$$

The above equations can be written as,

$$\frac{1}{\bar{\lambda}^2_{H|0}\bar{\lambda}^2_{H|1}} \sum_{i=1}^{n} \left\{ \frac{A_i}{p_1} Q_i - \frac{(1-A_i)}{1-p_1} R_i \right\} = 0$$

$$\frac{1}{\bar{\lambda}^2_{H|0}\bar{\lambda}^2_{H|1}} \sum_{i=1}^{n} A_i \left( \frac{1}{\lambda_{I|1}} - I_i \right) Q_i = 0$$

$$\frac{1}{\bar{\lambda}^2_{H|0}\bar{\lambda}^2_{H|1}} \sum_{i=1}^{n} A_i \left( \frac{1}{\lambda_{E|1}} - E_i \right) Q_i = 0$$

$$\frac{1}{\bar{\lambda}^2_{H|0}\bar{\lambda}^2_{H|1}} \sum_{i=1}^{n} A_i \left( \frac{1}{\lambda_{S|1}} - S_i \right) Q_i = 0$$

$$\sum_{i=1}^{n} A_i \left[ \frac{Q_i}{\bar{\lambda}^2_{H|0}\bar{\lambda}^2_{H|1}\lambda_{H|1}} - \frac{R_i}{\bar{\lambda}^2_{H|0}\bar{\lambda}^3_{H|1}} - \frac{2}{\bar{\lambda}^2_{H|0}\bar{\lambda}_{H|1}} - \frac{2\bar{p}_1}{\bar{\lambda}^2_{H|0}\bar{\lambda}_{H|1}} + \frac{6\bar{p}_1}{\bar{\lambda}^2_{H|1}\bar{\lambda}_{H|0}} + \frac{6-4\bar{p}_1}{\bar{\lambda}^3_{H|1}} \right] = 0$$

$$\frac{1}{\bar{\lambda}^2_{H|0}\bar{\lambda}^2_{H|1}} \sum_{i=1}^{n} \left\{ \frac{1-A_i}{p_0} R_i - \frac{A_i}{1-p_0} Q_i \right\} = 0$$

$$\frac{1}{\bar{\lambda}^2_{H|0}\bar{\lambda}^2_{H|1}} \sum_{i=1}^{n} (1-A_i) \left( \frac{1}{\lambda_{I|0}} - I_i \right) R_i = 0$$

$$\frac{1}{\bar{\lambda}^2_{H|0}\bar{\lambda}^2_{H|1}} \sum_{i=1}^{n} (1-A_i) \left( \frac{1}{\lambda_{E|0}} - E_i \right) R_i = 0$$

$$\frac{1}{\bar{\lambda}^2_{H|0}\bar{\lambda}^2_{H|1}} \sum_{i=1}^{n} (1-A_i) \left( \frac{1}{\lambda_{S|0}} - S_i \right) R_i = 0$$

$$\sum_{i=1}^{n} (1-A_i) \left[ \frac{R_i}{\bar{\lambda}^2_{H|0}\bar{\lambda}^2_{H|1}\lambda_{H|0}} - \frac{Q_i}{\bar{\lambda}^3_{H|0}\bar{\lambda}^2_{H|1}} - \frac{2}{\bar{\lambda}^2_{H|1}\bar{\lambda}_{H|0}} - \frac{2\bar{p}_0}{\bar{\lambda}^2_{H|1}\bar{\lambda}_{H|0}} + \frac{6\bar{p}_0}{\bar{\lambda}^2_{H|0}\bar{\lambda}_{H|1}} + \frac{6-4\bar{p}_0}{\bar{\lambda}^3_{H|0}} \right] = 0$$

Solving these newly obtained equations leads to the forms in Theorem 2.

___

**Appendix D: Evaluation Results with the Avatar Social Network**

Data on the avatar social network were collected from an online game company, including dialog, profile, and adoption data for 25,500 avatars registered in an online game over a one-year period. Dialog data contain 263.8 million dialog records among the 25,500 avatars; the profile of an avatar consists of 2 time-invariant attributes: gender and profession, and 11 time-variant behavioral attributes: accumulated online time, elapsed offline time since last log-out, shopping amount on dress, shopping amount on weapons, frequency of trading with other avatars, amount of money spent in trading, amount of money earned in trading, frequency of setting up a vendor stall, amount of money earned through vendor stalls, amount of money spent at other avatars' vendor stalls, and level[12]; and adoption data record week-by-week adoption of a virtual item, firstly available to purchase at the beginning of the one-year period, by the avatars, with an average weekly adoption rate of 0.40%. We created 52 snapshots of the avatar social network that include social entities ($v_i$) corresponding to avatars, the strength of a social tie ($x_{ij}^t$) measured as the average number of dialog records per week between avatars $v_i$ and $v_j$ by week $t$, and the intrinsic characteristics $\boldsymbol{c_i^t}$ consisting of avatar profile attributes. Following the evaluation procedure in §4.1, we compared our method with the benchmarks in Table 1. Similar to the mobile social network, we gathered additional data regarding weekly adoptions of all *other* virtual items by the same group of avatars during the study period for the implementation of the influence probability method. These virtual items are not necessities and firstly available to purchase during the study period. As shown in Table D1, the evaluation results are generally consistent with those of the mobile social network: (1) according to the Wilcoxon test, our method significantly outperforms each benchmark method ($p < 0.001$); (2) the AUCs of the cascade and the influence probability methods are marginally larger than that of random guess (0.5). These results provide additional evidence suggesting the predictive power of the proposed method and the importance of confounding factors in adoption probability predictions.

___

[12] An avatar can elevate its level through activities such as accomplishing a task.

| Evaluation Week ($T+1$) | LEMNB | CM1 | CM2 | CM3 | IP | NB | LWNB | SVM | k-NN |
|---|---|---|---|---|---|---|---|---|---|
| 3 | 0.5862 | 0.5031 | 0.5031 | 0.5031 | 0.5412 | 0.5533 | 0.5467 | 0.5735 | 0.5890 |
| 4 | 0.6498 | 0.5038 | 0.5038 | 0.5038 | 0.5022 | 0.6123 | 0.6023 | 0.6191 | 0.5826 |
| 5 | 0.6661 | 0.5034 | 0.5034 | 0.5034 | 0.5233 | 0.6375 | 0.6278 | 0.6190 | 0.6581 |
| 6 | 0.5720 | 0.5044 | 0.5044 | 0.5044 | 0.5311 | 0.5199 | 0.5110 | 0.5497 | 0.5540 |
| 7 | 0.6177 | 0.5231 | 0.5231 | 0.5231 | 0.5436 | 0.5592 | 0.5543 | 0.5069 | 0.5492 |
| 8 | 0.6412 | 0.5178 | 0.5178 | 0.5178 | 0.5347 | 0.5839 | 0.5781 | 0.5539 | 0.5618 |
| 9 | 0.6633 | 0.5089 | 0.5089 | 0.5089 | 0.5483 | 0.5076 | 0.5310 | 0.6330 | 0.6552 |
| 10 | 0.7528 | 0.5092 | 0.5092 | 0.5092 | 0.5209 | 0.5971 | 0.5734 | 0.6281 | 0.7077 |
| 11 | 0.7081 | 0.5097 | 0.5097 | 0.5097 | 0.5370 | 0.5439 | 0.5220 | 0.6415 | 0.6750 |
| 12 | 0.5772 | 0.5102 | 0.5102 | 0.5102 | 0.5422 | 0.5571 | 0.5243 | 0.5041 | 0.5701 |
| 13 | 0.7382 | 0.5086 | 0.5129 | 0.5089 | 0.5035 | 0.5737 | 0.6728 | 0.6061 | 0.6579 |
| 14 | 0.6234 | 0.5131 | 0.5134 | 0.5334 | 0.5363 | 0.5170 | 0.5072 | 0.5808 | 0.5810 |
| 15 | 0.6553 | 0.5204 | 0.5106 | 0.5141 | 0.5294 | 0.5153 | 0.5016 | 0.5876 | 0.6000 |
| 16 | 0.6750 | 0.5174 | 0.5171 | 0.5175 | 0.5263 | 0.5187 | 0.5037 | 0.5691 | 0.6572 |
| 17 | 0.6821 | 0.5114 | 0.5114 | 0.5114 | 0.5482 | 0.5323 | 0.5636 | 0.5977 | 0.6223 |
| 18 | 0.6896 | 0.5119 | 0.5119 | 0.5119 | 0.5373 | 0.5631 | 0.5328 | 0.6302 | 0.5321 |
| 19 | 0.6633 | 0.5122 | 0.5122 | 0.5122 | 0.5008 | 0.5951 | 0.5647 | 0.7120 | 0.6188 |
| 20 | 0.6267 | 0.5128 | 0.5128 | 0.5128 | 0.5375 | 0.5472 | 0.5388 | 0.6248 | 0.6084 |
| 21 | 0.6514 | 0.5133 | 0.5133 | 0.5133 | 0.5201 | 0.5766 | 0.5566 | 0.6678 | 0.6568 |
| 22 | 0.6975 | 0.5133 | 0.5133 | 0.5133 | 0.5000 | 0.5615 | 0.5041 | 0.6206 | 0.6733 |
| 23 | 0.6903 | 0.5136 | 0.5136 | 0.5136 | 0.5212 | 0.5659 | 0.5030 | 0.6665 | 0.6762 |
| 24 | 0.8316 | 0.5136 | 0.5136 | 0.5136 | 0.5000 | 0.6393 | 0.5478 | 0.6412 | 0.8003 |
| 25 | 0.8454 | 0.5139 | 0.5139 | 0.5139 | 0.5003 | 0.6398 | 0.5180 | 0.6357 | 0.7043 |
| 26 | 0.8253 | 0.5142 | 0.5142 | 0.5142 | 0.5003 | 0.6318 | 0.5610 | 0.6919 | 0.6669 |
| 27 | 0.7455 | 0.5142 | 0.5142 | 0.5142 | 0.5003 | 0.5338 | 0.5077 | 0.6789 | 0.7184 |
| 28 | 0.7592 | 0.5145 | 0.5145 | 0.5145 | 0.5003 | 0.5079 | 0.5330 | 0.6694 | 0.7396 |
| 29 | 0.8178 | 0.5155 | 0.5175 | 0.5158 | 0.5003 | 0.6258 | 0.5824 | 0.6846 | 0.7080 |
| 30 | 0.7542 | 0.5299 | 0.5313 | 0.5313 | 0.5000 | 0.5156 | 0.5619 | 0.7461 | 0.7302 |
| 31 | 0.8549 | 0.5151 | 0.5151 | 0.5151 | 0.5005 | 0.6603 | 0.6156 | 0.6910 | 0.7096 |
| 32 | 0.7843 | 0.5061 | 0.5127 | 0.5274 | 0.5013 | 0.5882 | 0.5439 | 0.7150 | 0.7318 |
| 33 | 0.7686 | 0.5156 | 0.5156 | 0.5156 | 0.5036 | 0.5154 | 0.5630 | 0.7576 | 0.7518 |
| 34 | 0.7371 | 0.5159 | 0.5159 | 0.5159 | 0.5052 | 0.5946 | 0.5459 | 0.7901 | 0.7384 |
| 35 | 0.6318 | 0.5167 | 0.5167 | 0.5167 | 0.5357 | 0.5685 | 0.5982 | 0.5686 | 0.5317 |
| 36 | 0.7799 | 0.5167 | 0.5167 | 0.5167 | 0.5338 | 0.6822 | 0.6102 | 0.6062 | 0.7832 |
| 37 | 0.7792 | 0.5175 | 0.5175 | 0.5175 | 0.5346 | 0.6772 | 0.6070 | 0.6249 | 0.7030 |
| 38 | 0.7889 | 0.5184 | 0.5184 | 0.5184 | 0.5034 | 0.5678 | 0.5226 | 0.7245 | 0.7568 |
| 39 | 0.6526 | 0.5189 | 0.5189 | 0.5189 | 0.5037 | 0.6372 | 0.5726 | 0.7445 | 0.6164 |
| 40 | 0.6791 | 0.5192 | 0.5192 | 0.5192 | 0.5032 | 0.5657 | 0.5047 | 0.6682 | 0.6344 |
| 41 | 0.7705 | 0.5200 | 0.5200 | 0.5200 | 0.5034 | 0.7143 | 0.5440 | 0.7220 | 0.7789 |
| 42 | 0.7346 | 0.5202 | 0.5202 | 0.5202 | 0.5034 | 0.6424 | 0.5557 | 0.6928 | 0.7450 |
| 43 | 0.6829 | 0.5203 | 0.5203 | 0.5203 | 0.5037 | 0.5971 | 0.5147 | 0.6527 | 0.6131 |
| 44 | 0.7834 | 0.5209 | 0.5209 | 0.5209 | 0.5037 | 0.6547 | 0.6443 | 0.6733 | 0.7231 |
| 45 | 0.8615 | 0.5214 | 0.5231 | 0.5214 | 0.5237 | 0.8567 | 0.8355 | 0.7243 | 0.8596 |
| 46 | 0.7720 | 0.5146 | 0.5027 | 0.5133 | 0.5037 | 0.6510 | 0.6213 | 0.6859 | 0.6940 |
| 47 | 0.7254 | 0.5188 | 0.5355 | 0.5355 | 0.5037 | 0.7213 | 0.6830 | 0.5895 | 0.7097 |
| 48 | 0.7033 | 0.5176 | 0.5144 | 0.5144 | 0.5042 | 0.6789 | 0.6964 | 0.6296 | 0.7325 |
| 49 | 0.7340 | 0.5136 | 0.5136 | 0.5124 | 0.5283 | 0.6606 | 0.7032 | 0.5949 | 0.7020 |
| 50 | 0.7889 | 0.5245 | 0.5245 | 0.5245 | 0.5378 | 0.7741 | 0.7818 | 0.6995 | 0.7811 |
| 51 | 0.8537 | 0.5280 | 0.5290 | 0.5290 | 0.5000 | 0.8548 | 0.8553 | 0.8333 | 0.8652 |
| 52 | 0.6856 | 0.5276 | 0.5276 | 0.5276 | 0.5000 | 0.6141 | 0.6930 | 0.6031 | 0.6807 |
| Avg. | 0.7192 | 0.5153 | 0.5155 | 0.5163 | 0.5165 | 0.6062 | 0.5829 | 0.6486 | 0.6779 |
| Std. | 0.0761 | 0.0060 | 0.0069 | 0.0072 | 0.0169 | 0.0804 | 0.0826 | 0.0686 | 0.0799 |

**Table D1:** Comparative AUC Analyses of Proposed and Benchmark Methods

**Appendix E: AUC Diagrams: Proposed Method versus Benchmark Methods**

Using data in Table 2, we plot AUCs of the proposed method and each benchmark method across 50 evaluation weeks. According to our discussion in §4.2, the performance of learning-based predictive methods (i.e., LEMNB, NB, LWNB, SVM, and k-NN) vary across evaluation weeks and the performance of cascade methods (i.e., CM1, CM2, CM3, and IP) is relatively stable, as reflected in the following figures. Furthermore, the AUCs of different learning-based predictive methods may vary differently across evaluation weeks because of the differences in the patterns they learned.

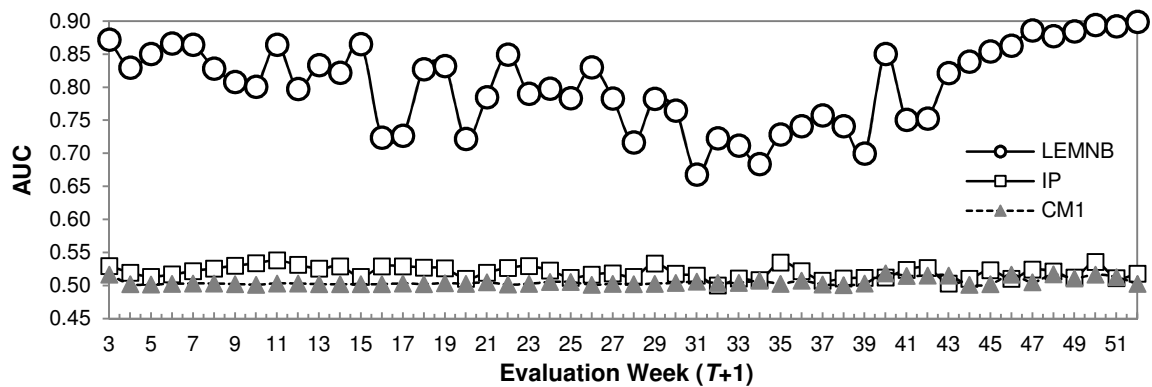

**Figure E1:** AUC Across Evaluation Weeks: LEMNB vs. IP vs. CM1

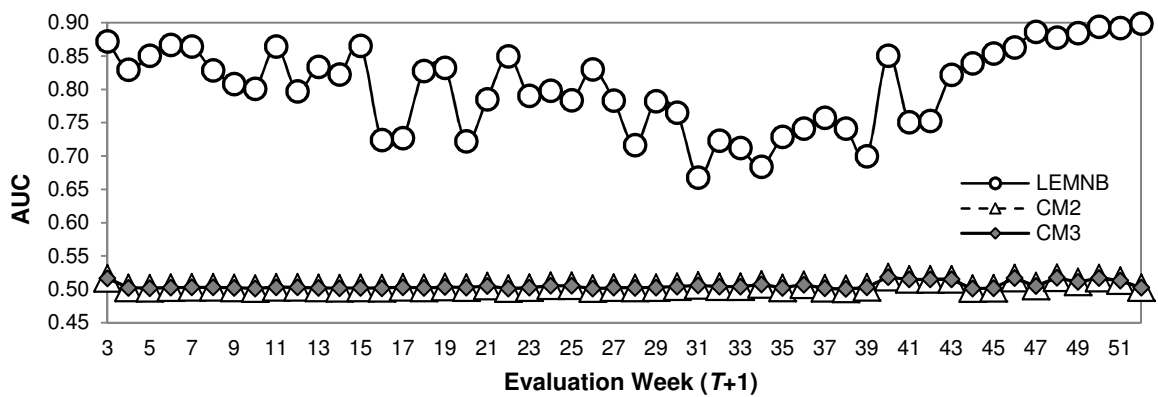

**Figure E2:** AUC Across Evaluation Weeks: LEMNB vs. CM2 vs. CM3

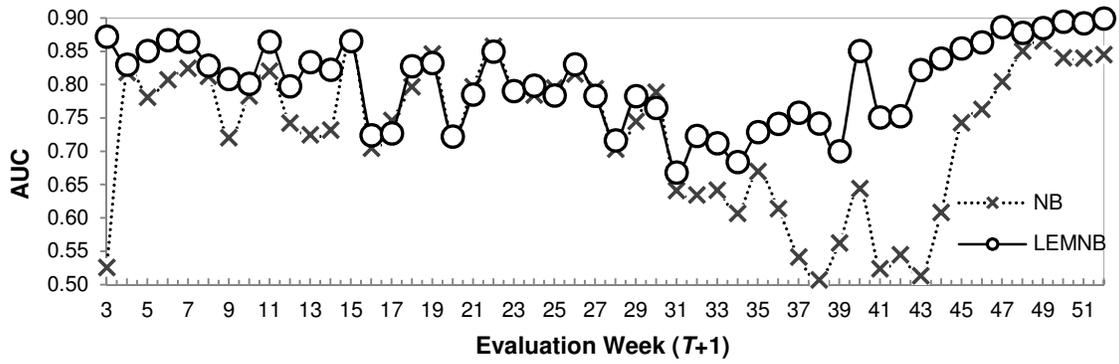

**Figure E3:** AUC Across Evaluation Weeks: LEMNB vs. NB

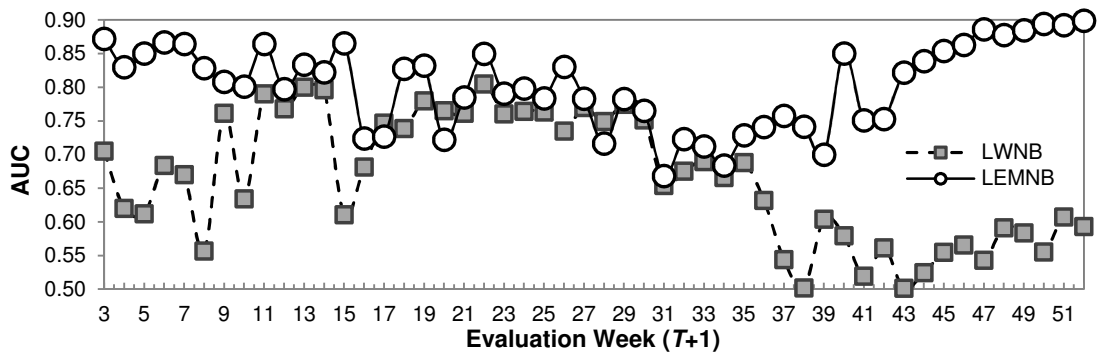

**Figure E4:** AUC Across Evaluation Weeks: LEMNB vs. LWNB

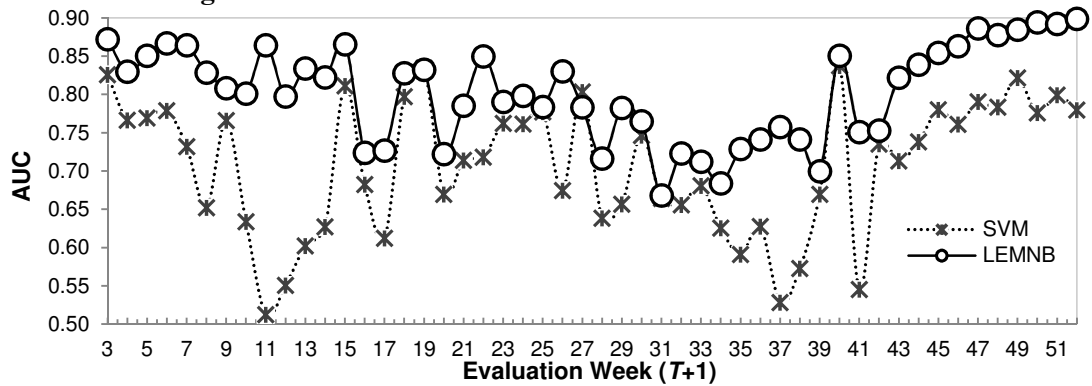

**Figure E5:** AUC Across Evaluation Weeks: LEMNB vs. SVM

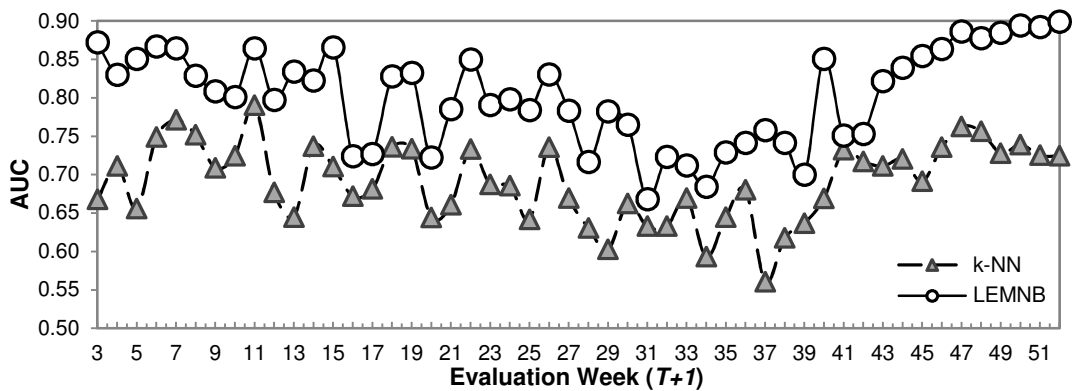

**Figure E6:** AUC Across Evaluation Weeks: LEMNB vs. k-NN

**Appendix F: Extending Proposed Method with Additional Factor: A Preliminary Analysis**

For a social entity $v$, with a set $F$ of adopter neighbors, let $e(F)$ denote the number of social ties connecting a pair of adopter neighbors in $F$. According to (Backstrom et al. 2006), the connectedness $Z(v)$ of $v$'s adopter neighbors can be measured as

$$Z(v) = \frac{e(F)}{\binom{|F|}{2}} \tag{G1}$$

where $\binom{|F|}{2}$ is the number of all possible social ties connecting adopter neighbors in $F$. To consider the connectedness factor, we have to reconstruct training and test data by calculating $Z(v)$ for each training or test record. More importantly, we need to adjust our method to predict a social entity's adoption probability, on the basis of influence, equivalence, similarity, and hidden confounding powers on the entity as well as the connectedness of the entity's adopter neighbors.

Theorems 1 and 2 still hold when considering the connectedness factor, because of the Naïve Bayes's conditional independence among factors. Therefore, we can estimate parameters for the connectedness factor in the same way as we estimate parameters for other factors (e.g., influence power). Specifically, we can initialize parameters $\bar{\lambda}_{Z|1}$ and $\bar{\lambda}_{Z|0}$ for the connectedness factor as follows, similar to parameter initialization for influence power in equations (33) and (34):

$$\bar{\lambda}_{Z|1} = \frac{\sum_{i=1}^{n} A_i}{\sum_{i=1}^{n} A_i Z_i}, \tag{G2}$$

$$\bar{\lambda}_{Z|0} = \frac{\sum_{i=1}^{n} (1-A_i)}{\sum_{i=1}^{n} (1-A_i) Z_i}, \tag{G3}$$

where $Z_i$ denotes the connectedness of adopter neighbors for training record $i$. Furthermore, we can update parameters $\hat{\lambda}_{Z|1}$ and $\hat{\lambda}_{Z|0}$ as follows, similar to parameter updates for influence power in equations (23) and (24):

$$\hat{\lambda}_{Z|1} = \frac{\sum_{i=1}^{n} A_i Q_i}{\sum_{i=1}^{n} A_i Z_i Q_i}, \tag{G4}$$

$$\hat{\lambda}_{Z|0} = \frac{\sum_{i=1}^{n} (1-A_i) R_i}{\sum_{i=1}^{n} (1-A_i) Z_i R_i}. \tag{G5}$$

Consideration of the connectedness factor requires several adjustments of the proposed method: (1) initializing the parameters for the connectedness factor with equations (G2) and (G3) in parameter initialization; and (2) updating parameters for the connectedness factor with equations (G4) and (G5) in parameter update. For clarity, we now name this adjusted method, which considers the additional connectedness factor, LEMNB+.

In Table F1, we compare the performance of LEMNB+ and LEMNB (i.e., the original method without considering the connectedness factor) using both social networks we studied. We apply the Wilcoxon test (Demsar 2006) to the AUCs in Table F1; statistical testing results suggest that, for each social network, LEMNB+ significantly outperforms LEMNB ($p < 0.01$). Our evaluation results provide evidence suggesting that the connectedness of adopter neighbors could be an important predictor of adoption behaviors, consistent with Backstrom et al. (2006). Further, we demonstrate a viable way to incorporate additional factors into the proposed method through this analysis.

| Evaluation Week | Mobile Social Network | | Avatar Social Network | |
|---|---|---|---|---|
| ($T$+1) | LEMNB | LEMNB+ | LEMNB | LEMNB+ |
| 3 | 0.8722 | 0.8955 | 0.5862 | 0.6127 |
| 4 | 0.8299 | 0.8741 | 0.6498 | 0.6637 |
| 5 | 0.8505 | 0.9010 | 0.6661 | 0.6988 |
| 6 | 0.8666 | 0.9086 | 0.5720 | 0.5728 |
| 7 | 0.8643 | 0.9036 | 0.6177 | 0.6672 |
| 8 | 0.8286 | 0.7946 | 0.6412 | 0.6947 |
| 9 | 0.8083 | 0.7808 | 0.6633 | 0.6633 |
| 10 | 0.8011 | 0.7659 | 0.7528 | 0.7528 |
| 11 | 0.8643 | 0.8346 | 0.7081 | 0.7275 |
| 12 | 0.7974 | 0.8458 | 0.5772 | 0.5602 |
| 13 | 0.8339 | 0.8777 | 0.7382 | 0.7619 |
| 14 | 0.8223 | 0.7947 | 0.6234 | 0.5689 |
| 15 | 0.8654 | 0.8567 | 0.6553 | 0.6553 |
| 16 | 0.7240 | 0.7041 | 0.6750 | 0.6750 |
| 17 | 0.7270 | 0.7399 | 0.6821 | 0.6821 |
| 18 | 0.8276 | 0.8632 | 0.6896 | 0.6448 |
| 19 | 0.8323 | 0.8822 | 0.6633 | 0.6633 |
| 20 | 0.7221 | 0.7603 | 0.6267 | 0.6296 |
| 21 | 0.7852 | 0.7568 | 0.6514 | 0.6514 |
| 22 | 0.8497 | 0.8935 | 0.6975 | 0.6769 |
| 23 | 0.7905 | 0.7628 | 0.6903 | 0.6809 |
| 24 | 0.7982 | 0.8363 | 0.8316 | 0.8596 |
| 25 | 0.7838 | 0.7627 | 0.8454 | 0.8595 |
| 26 | 0.8302 | 0.7993 | 0.8253 | 0.8253 |
| 27 | 0.7832 | 0.7672 | 0.7455 | 0.7658 |
| 28 | 0.7165 | 0.6872 | 0.7592 | 0.7529 |
| 29 | 0.7825 | 0.8099 | 0.8178 | 0.8178 |
| 30 | 0.7650 | 0.7569 | 0.7542 | 0.7542 |
| 31 | 0.6681 | 0.7023 | 0.8549 | 0.8655 |
| 32 | 0.7232 | 0.7459 | 0.7843 | 0.7843 |
| 33 | 0.7121 | 0.7516 | 0.7686 | 0.7686 |
| 34 | 0.6840 | 0.6987 | 0.7371 | 0.8544 |
| 35 | 0.7289 | 0.7557 | 0.6318 | 0.6065 |
| 36 | 0.7413 | 0.7645 | 0.7799 | 0.7799 |
| 37 | 0.7577 | 0.7608 | 0.7792 | 0.7792 |
| 38 | 0.7414 | 0.7439 | 0.7889 | 0.8813 |
| 39 | 0.6999 | 0.7030 | 0.6526 | 0.8010 |
| 40 | 0.8505 | 0.8838 | 0.6791 | 0.7503 |
| 41 | 0.7509 | 0.7570 | 0.7705 | 0.8006 |
| 42 | 0.7528 | 0.7575 | 0.7346 | 0.7581 |
| 43 | 0.8218 | 0.8225 | 0.6829 | 0.7317 |
| 44 | 0.8392 | 0.8399 | 0.7834 | 0.7834 |
| 45 | 0.8542 | 0.8563 | 0.8615 | 0.8228 |
| 46 | 0.8633 | 0.8649 | 0.7720 | 0.8013 |
| 47 | 0.8864 | 0.8862 | 0.7254 | 0.7249 |
| 48 | 0.8777 | 0.8777 | 0.7033 | 0.6880 |
| 49 | 0.8847 | 0.8847 | 0.7340 | 0.7058 |
| 50 | 0.8942 | 0.8948 | 0.7889 | 0.8005 |
| 51 | 0.8922 | 0.8933 | 0.8537 | 0.8564 |
| 52 | 0.8990 | 0.8990 | 0.6856 | 0.7225 |
| Avg. | 0.8029 | 0.8112 | 0.7192 | 0.7321 |
| Std. | 0.0625 | 0.0672 | 0.0761 | 0.0832 |

**Table F1:** Comparative AUC Analyses of LEMNB and LEMNB+

**Appendix G: Evaluating Proposed Method with an Alternative Operationalization of Entity Similarity**

We examine whether an alternative operationalization of entity similarity could affect the performance of our method. Specifically, the distance function by Aha et al. (1991) can be used to measure the similarity between entities with a mix of nominal, real-valued, and integer attributes. Like the distance function used by our method, the function by Aha et al. (1991) still computes entity distance on each feature with equation (44) or (45) but integrates entity distance on each feature differently. In particular, the distance between entities $x$ and $y$ is computed as (Aha et al. 1991),

$$d(x, y) = \sqrt{\sum_{i=1}^{n} d(x_i, y_i)^2}, \tag{G1}$$

where $d(x_i, y_i)$ is computed using equation (44) or (45). In Table G1, we compare the performance between our method LEMNB and a variance using the distance function by Aha et al. (1991), namely LEMNBA. We apply the Wilcoxon test (Demsar 2006) to the AUCs across 50 evaluations in Table G1; our testing results suggest that, for each social network we studied, the performance difference between LEMNB and LEMNBA is not significant ($p > 0.1$), thereby suggesting the robustness of our method across these commonly used distance functions.

| Evaluation Week | Mobile Social Network | | Avatar Social Network | |
|---|---|---|---|---|
| (*T*+1) | LEMNB | LEMNBA | LEMNB | LEMNBA |
| 3 | 0.8722 | 0.8870 | 0.5862 | 0.5605 |
| 4 | 0.8299 | 0.8393 | 0.6498 | 0.5861 |
| 5 | 0.8505 | 0.8608 | 0.6661 | 0.6182 |
| 6 | 0.8666 | 0.8772 | 0.5720 | 0.5437 |
| 7 | 0.8643 | 0.8729 | 0.6177 | 0.6483 |
| 8 | 0.8286 | 0.8274 | 0.6412 | 0.7088 |
| 9 | 0.8083 | 0.8197 | 0.6633 | 0.6624 |
| 10 | 0.8011 | 0.7836 | 0.7528 | 0.6960 |
| 11 | 0.8643 | 0.8732 | 0.7081 | 0.7438 |
| 12 | 0.7974 | 0.7902 | 0.5772 | 0.5731 |
| 13 | 0.8339 | 0.8529 | 0.7382 | 0.7010 |
| 14 | 0.8223 | 0.8390 | 0.6234 | 0.5786 |
| 15 | 0.8654 | 0.8742 | 0.6553 | 0.7098 |
| 16 | 0.7240 | 0.7357 | 0.6750 | 0.6219 |
| 17 | 0.7270 | 0.7238 | 0.6821 | 0.7094 |
| 18 | 0.8276 | 0.8351 | 0.6896 | 0.6541 |
| 19 | 0.8323 | 0.8402 | 0.6633 | 0.6463 |
| 20 | 0.7221 | 0.7323 | 0.6267 | 0.6320 |
| 21 | 0.7852 | 0.7944 | 0.6514 | 0.6790 |
| 22 | 0.8497 | 0.8763 | 0.6975 | 0.6701 |
| 23 | 0.7905 | 0.8046 | 0.6903 | 0.6846 |
| 24 | 0.7982 | 0.7940 | 0.8316 | 0.8241 |
| 25 | 0.7838 | 0.7907 | 0.8454 | 0.8794 |
| 26 | 0.8302 | 0.8387 | 0.8253 | 0.8930 |
| 27 | 0.7832 | 0.7999 | 0.7455 | 0.7753 |
| 28 | 0.7165 | 0.7148 | 0.7592 | 0.7435 |
| 29 | 0.7825 | 0.7798 | 0.8178 | 0.7591 |
| 30 | 0.7650 | 0.7759 | 0.7542 | 0.7843 |
| 31 | 0.6681 | 0.6679 | 0.8549 | 0.8046 |
| 32 | 0.7232 | 0.6861 | 0.7843 | 0.8489 |
| 33 | 0.7121 | 0.6783 | 0.7686 | 0.7993 |
| 34 | 0.6840 | 0.6700 | 0.7371 | 0.7340 |
| 35 | 0.7289 | 0.7538 | 0.6318 | 0.6380 |
| 36 | 0.7413 | 0.7350 | 0.7799 | 0.7606 |
| 37 | 0.7577 | 0.7120 | 0.7792 | 0.7407 |
| 38 | 0.7414 | 0.6800 | 0.7889 | 0.8204 |
| 39 | 0.6999 | 0.6711 | 0.6526 | 0.6657 |
| 40 | 0.8505 | 0.8580 | 0.6791 | 0.7313 |
| 41 | 0.7509 | 0.7188 | 0.7705 | 0.8008 |
| 42 | 0.7528 | 0.7167 | 0.7346 | 0.7437 |
| 43 | 0.8218 | 0.8097 | 0.6829 | 0.6799 |
| 44 | 0.8392 | 0.8557 | 0.7834 | 0.8090 |
| 45 | 0.8542 | 0.8261 | 0.8615 | 0.8321 |
| 46 | 0.8633 | 0.8506 | 0.7720 | 0.7469 |
| 47 | 0.8864 | 0.8428 | 0.7254 | 0.7254 |
| 48 | 0.8777 | 0.8371 | 0.7033 | 0.6454 |
| 49 | 0.8847 | 0.8702 | 0.7340 | 0.7115 |
| 50 | 0.8942 | 0.8576 | 0.7889 | 0.7842 |
| 51 | 0.8922 | 0.8711 | 0.8537 | 0.8878 |
| 52 | 0.8990 | 0.8674 | 0.6856 | 0.7021 |
| **Avg.** | 0.8029 | 0.7974 | 0.7192 | 0.7180 |
| **Std.** | 0.0625 | 0.0678 | 0.0761 | 0.0863 |

**Table G1:** Comparative AUC Analyses of LEMNB and LEMNBA